\documentclass[nonacm,sigplan]{acmart}

\makeatletter
\def\@ACM@checkaffil{
    \if@ACM@instpresent\else
    \ClassWarningNoLine{\@classname}{No institution present for an affiliation}%
    \fi
    \if@ACM@citypresent\else
    \ClassWarningNoLine{\@classname}{No city present for an affiliation}%
    \fi
    \if@ACM@countrypresent\else
        \ClassWarningNoLine{\@classname}{No country present for an affiliation}%
    \fi
}
\makeatother

\usepackage{soul}
\usepackage{graphicx}
\usepackage{subcaption}
\usepackage[]{hyperref}
\usepackage{balance}
\begin{document}

\title{Exposing Shadow Branches}

\author{Chrysanthos Pepi\textsuperscript{\symbol{42}}}
\affiliation{
 \institution{Texas A\&M University}}
\email{cpepis@tamu.edu}

\author{Bhargav Reddy Godala\textsuperscript{\symbol{42}}}
\affiliation{
 \institution{Princeton University}}
\email{bgodala@princeton.edu}

\author{Krishnam Tibrewala}
\affiliation{
 \institution{Texas A\&M University}}
\email{krishnamtibrewala@tamu.edu}

\author{Gino A. Chacon}
\affiliation{
 \institution{Intel Corporation}}
\email{ginoachacon@intel.com}

\author{Paul V. Gratz}
\affiliation{
 \institution{Texas A\&M University}}
\email{pgratz@gratz1.com}

\author{Daniel A. Jiménez}
\affiliation{
 \institution{Texas A\&M University}}
\email{djimenez@tamu.edu}

\author{Gilles A. Pokam}
\affiliation{
 \institution{Intel Corporation}}
\email{gilles.a.pokam@intel.com}

\author{David I. August}
\affiliation{
 \institution{Princeton University}}
\email{august@princeton.edu}

\thanks{
\symbol{42} Authors have contributed equally.
}

\begin{abstract}
Modern processors implement a decoupled front-end, often using a form of Fetch Directed Instruction Prefetching (FDIP), to avoid front-end stalls.  FDIP is driven by the Branch Prediction Unit (BPU), relying on the BPU's accuracy and branch target tracking structures to speculatively fetch instructions into the Instruction Cache (L1-I cache).  As contemporary data center applications become more complex, their code footprints also grow, resulting in a high number of Branch Target Buffer (BTB) misses. These BTB missing branches typically have previously been decoded and placed in the BTB, but have since been evicted, leading to BTB misses now. FDIP can alleviate L1-I cache misses, but its reliance on the BPU's tracking structures means that when it encounters a BTB miss, the BPU may not identify the current instruction as a branch to FDIP.  This can prevent FDIP from prefetching or cause it to speculate down the wrong path, further polluting the L1-I cache.

We observe that the vast majority, 75\%, of BTB-missing, unidentified branches are actually present in instruction cache lines that FDIP has previously fetched.  Nevertheless, these missing branches have not yet been decoded and inserted into the BTB.  This is because the instruction line is decoded from an entry point (which is the target of the previous taken branch) till an exit point (taken branch). We call branch instructions present in the ignored portion of the cache line ``Shadow Branches.''  
Here we present Skia, a novel shadow branch decoding technique that identifies and decodes unused bytes in cache lines fetched by FDIP, inserting them into a Shadow Branch Buffer (SBB). The SBB is accessed in parallel with the BTB, allowing FDIP to speculate despite a BTB miss.

With a minimal storage state of 12.25KB, Skia delivers a geomean speedup of $\sim$5.7\% over an 8K-entry BTB (78KB) and $\sim$2\% versus adding an equal amount of state to the BTB, across 16 front-end bound applications.  Since many branches stored in the SBB are distinct compared to those in a similarly sized BTB, we consistently observe greater performance gains with Skia across all examined sizes until saturation.
\end{abstract}

\maketitle 
\pagestyle{plain} 

\section{Introduction}
\label{sec:introduction}
Modern data center and commercial workloads place enormous pressure on the processor core front-end and instruction cache~\cite{ayers2019asmdb, kanev2015profiling}.  Many recent works explore mechanisms to reduce front-end pressure and instruction cache misses~\cite{kumar2017boomerang, kaynak2015confluence, khan2021twig, song2022thermometer, kumar2018blasting, ansari2020divide, ros2023wrong, godala2024pdip, fdip, ripple, bolt, khan2022whisper, schall2022lukewarm, khan2020spy, kolli2013rdip}.  Instruction prefetching has been explored extensively to improve the efficiency of the instruction cache~\cite{godala2024pdip, ros2023wrong, ansari2020divide, kumar2018blasting, kaynak2015confluence, kumar2017boomerang, fdip, schall2022lukewarm, kolli2013rdip}. 
 In particular, Fetch Directed Instruction Prefetching (FDIP)~\cite{fdip} is a common front-end design that decouples the instruction fetch and branch prediction structures, allowing for an instruction prefetcher to runahead of the current instruction stream.  A decoupled front-end alleviates stalls due to instruction cache misses however its predictions are solely dependent on the branch predictor to provide instruction addresses.

\begin{figure}[t]
  \centering
  \includegraphics[width=\columnwidth]{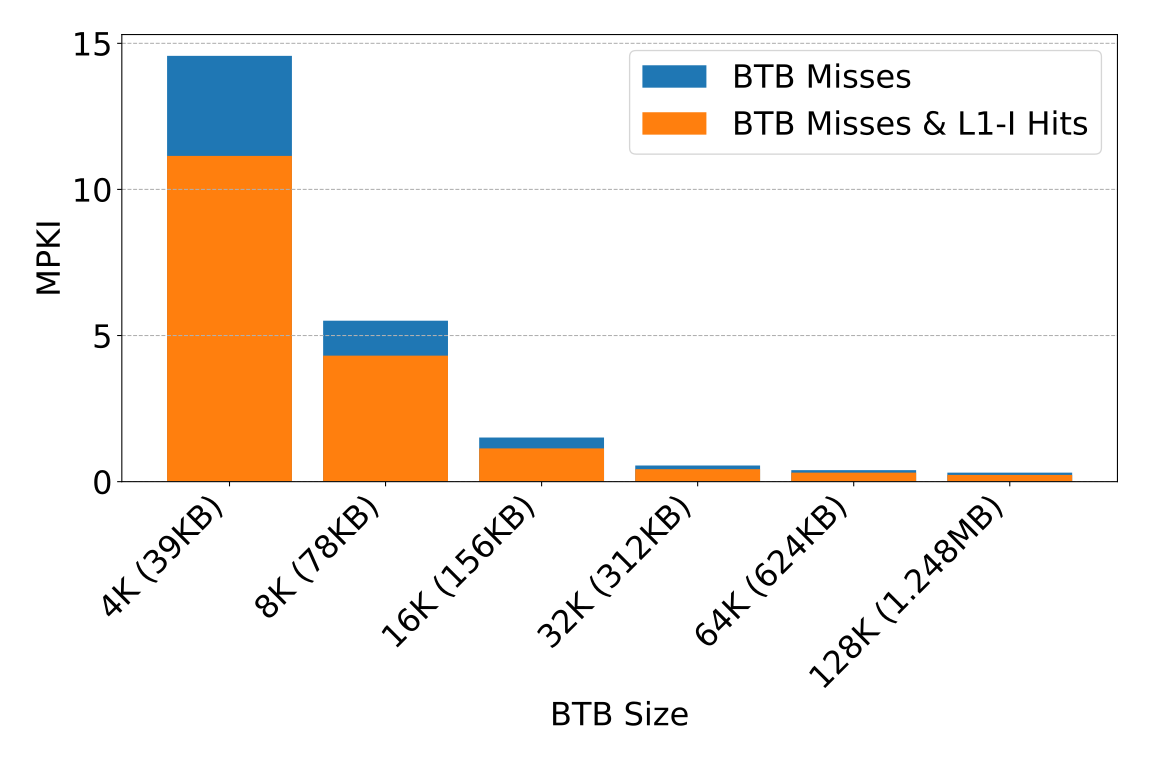}
  \caption{BTB misses and the proportion of these misses that coincide with L1-I hits across different BTB sizes.}
  \label{fig:btb_misses_per_config}
\end{figure}

\begin{figure}[h]
  \centering
  \includegraphics[width=\columnwidth]{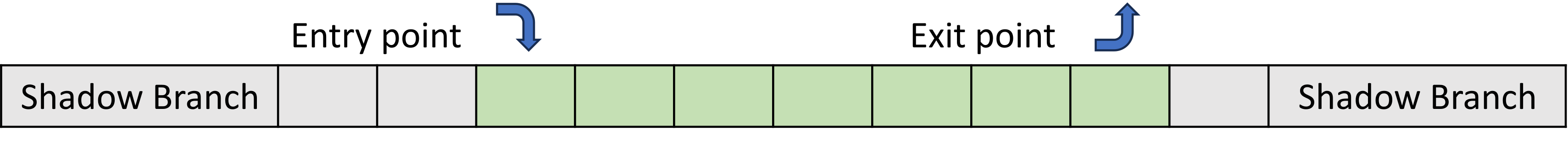}
  \caption{A cache line indicating the presence of branches both before and after the entry/exit point.}
  \label{fig:abs_head_tail}
\end{figure}

Recent works demonstrate that the front-end is a considerable source of performance loss~\cite{ayers2019asmdb}, with upwards of 53\% of performance~\cite{godala2024pdip} bounded by the front-end.  Figure~\ref{fig:btb_misses_per_config} shows the average misses per kilo-instructions (MPKI) across a set of 16 L1-I bound commercial workloads.  Each bar shows the total MPKI, and the MPKI of BTB misses where the branch is already in the L1-I cache is shown in orange.  As the figure shows, the vast majority, 75\%, for a nominally sized, 8K-entry BTB, of BTB-missing, unidentified branches are actually present in instruction cache lines that FDIP has previously fetched.  To the best of our knowledge, we are the first to observe and recognize this behavior. 
 Nevertheless, these BTB-missing branches have yet to be decoded and inserted into the BTB.  This is because they lie either before the branch target that brought the line into the cache or after a taken branch that leaves the cache line.  We call these branches in the shadow of the executed basic block "Shadow Branches" as pictured in Figure~\ref{fig:abs_head_tail}.

These missing shadow branches are typically less frequently encountered, or "cold" branches. We define "cold" branches as those that, while they recur frequently throughout the program's execution, are separated by a significant number of other branches between recurrences. The total number of branches in the program is large enough that these cold branches are typically evicted from the BTB before they can be seen again. As we define them, these "cold" branch misses are a form of capacity miss, not a compulsory miss. Compulsory misses occur only the first time a branch is encountered, whereas cold branches may cause many misses due to the large footprint of branches in these programs.

Nevertheless, these cold shadow branches contribute significantly to branch resteers. However, we find that these cold branches are typically present in the instruction cache but are not decoded or added to the BTB. This occurs because they are not part of the executing basic block that loads the cache line into the L1-I cache. A good example of code that exhibits this behavior is when frequently used functions are placed next to less frequently used, colder functions in the binary. While the branches and returns in frequently used functions would be correctly retained in the BTB, the less used functions—co-located on the same cache line—are never decoded until they are finally executed later, which leads to a BTB miss on an L1 instruction cache hit.

A commonly proposed solution is to enhance FDIP to address both L1-I and BTB misses~\cite{kumar2018blasting, kaynak2015confluence, kumar2017boomerang}.  These prefetchers use the information from the BTB to run ahead of the current instruction stream and proactively fill the L1-I.  While these prefetchers can reduce the BTB miss rate, they are ultimately dependent on the contents of the BTB to generate predictions, with cold branches unlikely to be on their predicted path.  Any mispredictions due to prefetching down the speculative path risks polluting the L1-I and BTB, further exacerbating the front-end bottleneck.  Moreover, most of these approaches operate on fixed-length instruction sets since the predecoder relies on the perfect, 4-byte alignment of all instructions.  Few proposals~\cite{ansari2020divide} address the challenges of variable-length instruction sets and those that do similarly seek to fill the BTB based on L1-I misses requiring virtualized metadata storage in the Last-Level Cache.

We note that FDIP, in the process of fetching and forwarding cache lines containing code to be executed to the Fetch Engine to forward to decode, often also forwards the bytes containing shadow branches co-resident with true-path branches on those cache lines.  Thus, these shadow branches are already fetched into the front-end of the machine.  We will leverage this fact to decode these shadow branches in parallel with the true-path branches.

\begin{figure}[t]
    \centering
    \includegraphics[width=\columnwidth]{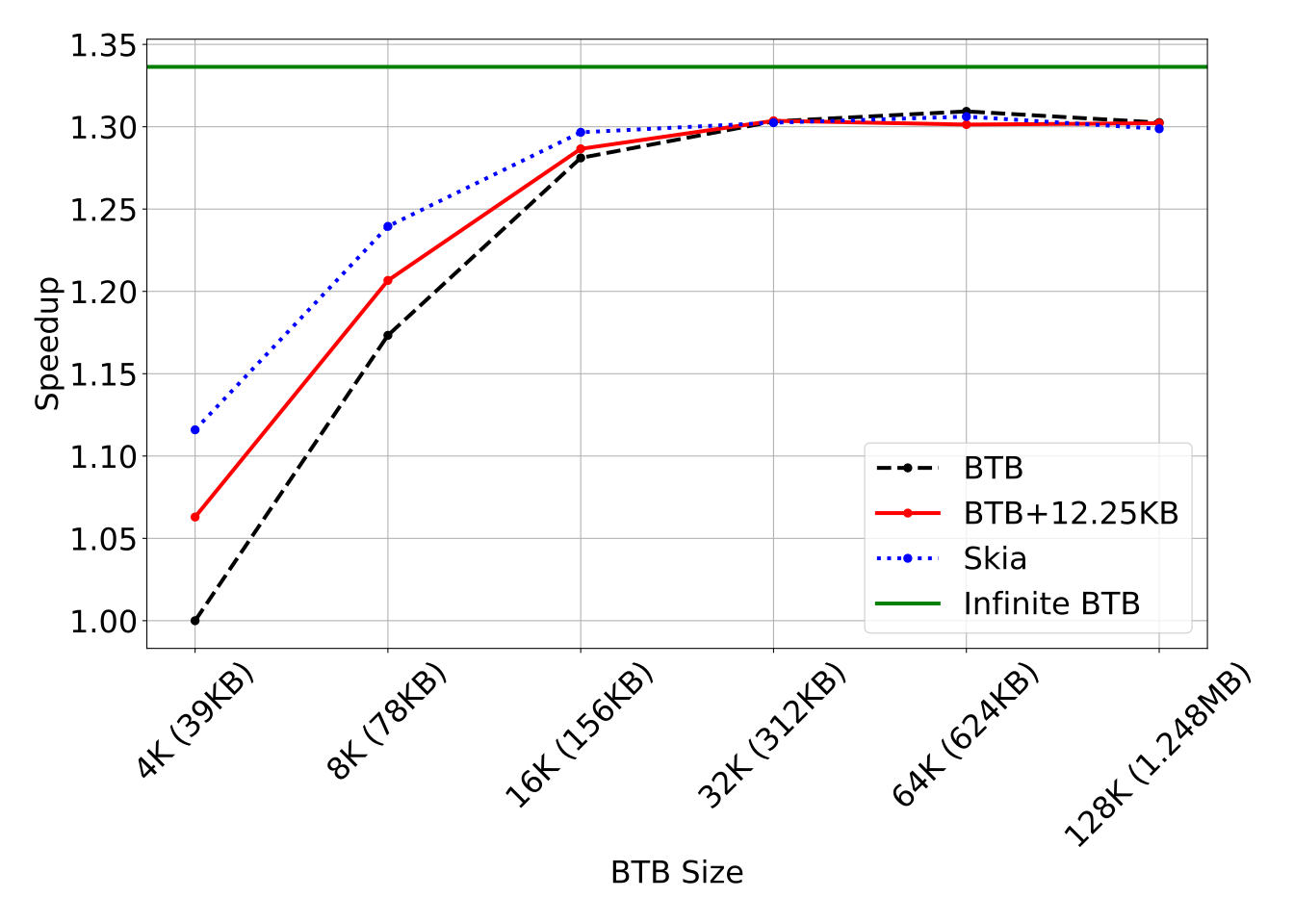}
    \caption{Speedup (Geomean of all benchmarks examined) relative to different BTB sizes. Four configurations are compared: BTB, BTB+12.25KB, a BTB with our 12.25KB SBB (BTB+SBB), and Infinite, Fully Associative BTB.}
    \label{fig:ipc}
\end{figure}

\subsection{Key Contributions}
To address the high BTB miss rate in contemporary commercial workloads, here we introduce Skia~\footnote{Skia is Greek for "shadow", thus we use this term to signify our technique.}, a new and novel technique to leverage the previously not-decoded bytes on FDIP fetched cache lines.  

Figure~\ref{fig:ipc} summarizes the performance benefit that Skia can provide across a range of baseline BTB sizes.  In the figure we show the speedup of several designs, normalized against a small, 4K entry BTB.  The performance here is the geomean average across all benchmarks examined. In the plot, the lowest black dashed line represents the performance of a standard BTB with the specified number of entries. The red solid line indicates the performance of a standard BTB with an additional 12.25KB of storage, matching the size of the default Skia structure (ISO hardware budget given to BTB). The blue dotted line shows the performance of a system enhanced with Skia. 
The green line at the top represents the performance of a system with an infinite, fully associative BTB.
Across all BTB sizes, our proposed Skia design consistently achieves nearly twice the performance improvement (until saturation), versus adding the same amount of storage to the BTB. This demonstrates Skia’s significant effectiveness in enhancing system performance.

This paper makes the following contributions:
\begin{itemize}
    \item The first work to observe that the majority of BTB missing branches, reside in the L1-I cache and are available for decode without requiring a cache fill.
    \item Introduces Skia, a new and novel mechanism designed for the speculative identification and decoding of shadow branches, applicable to both fixed and variable-length instructions.
    \item Introduces an efficient and small structure, the Shadow Branch Buffer (SBB), for storing these shadow branches that can be filled off the critical path and accessed in parallel with the real BTB, achieving a geomean $\sim$5.7\% performance improvement with only 12.25KB of state.
    \item The branches stored in the SBB are notably different compared to those in a similarly sized BTB.  Our findings indicate that allocating the same amount of state storage space exclusively to the SBB, as opposed to BTB, results in higher performance gains across nearly all BTB sizes.
\end{itemize}

\section{Background and Motivation}
\label{sec:background}
This section reviews the background of modern processor front-ends, the decoding of CISC variable-length instructions, and branch types. It also discusses the motivation for the work concerning BTB scaling and shadow branch placement.

\subsection{Fetch-Directed Instruction Prefetching}
\begin{figure}[h]
    \centering
    \includegraphics[width=\columnwidth]{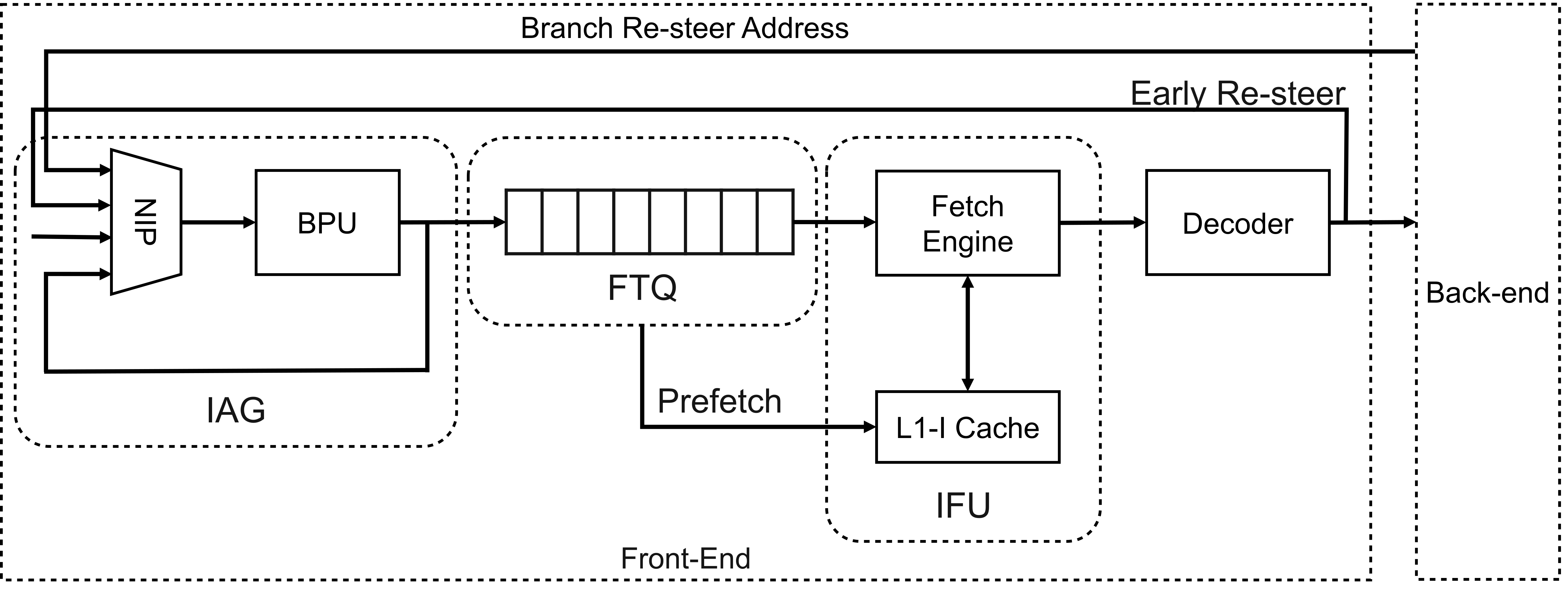}
    \caption{Generic decoupled front-end microarchitecture.}
    \label{fig:FDIP}
\end{figure}

Figure~\ref{fig:FDIP} depicts the typical superscalar core front-end microarchitecture.  Notably, it shows the decoupling of the Instruction Fetch Unit (IFU) from the Instruction Address Generator (IAG) by the presence of the Fetch Target Queue (FTQ).  The Branch Prediction Unit (BPU), a critical component of the IAG, comprises the conditional branch predictor, the direct jump address predictor (commonly referred to as the Branch Target Buffer or BTB), the indirect jump predictor, and a return address stack.  These elements collectively provide prediction data for the IAG, enabling it to speculatively compute the address of the next Basic Block and insert this in the form of predicted cache line addresses into the FTQ.

The FTQ operates as a First-In-First-Out queue, capturing the targets computed by the IAG along the predicted execution path.  Each entry in the FTQ corresponds to a Basic Block, and the cache lines associated with each Basic Block are identified and prefetched into the L1-I cache.  This prefetching mechanism allows non-resident instruction blocks to be prefetched into the L1-I cache upon entry into the FTQ, rather than waiting for them to be fetched on demand when the address reaches the IFU.  Typically, the IAG operates ahead of the back-end, ensuring that the FTQ remains consistently filled, except in cases of pipeline squashes.

The effectiveness of FDIP hinges greatly on the BTB.  When FDIP encounters a BTB miss, it risks failing to correctly identify the next branch, potentially leading to fetching incorrect L1-I cache lines for taken branches.  These unused prefetches are placed in the cache regardless of the later branch resteer, leading to pollution of the L1-I.  As we observe in Figure~\ref{fig:btb_misses_per_config}, with a high number of BTB missing branches being in the shadow bytes of already fetched cache lines, the opportunity exists to improve FDIP effectiveness by decoding as many of them as possible.

\subsection{CISC Variable Length Instruction Decoding}
Typical CISC (Complex Instruction Set Computer) instruction sets, such as x86, encode instructions using a variable number of bytes.  As a result, decoding CISC instructions incurs some serialization, making parallel decoding of multiple instructions, as required in superscalar cores, more challenging since the beginning byte of an instruction is only known in two cases: one, when the prior instruction is decoded and its length is known; and two, when a branch target, \emph{i.e.} entry point to the cache line, location is known.

In CISC instruction decoding, these variable-length instructions must be parsed and translated into micro-operations (uops) that the processor can execute.  Decoding begins when the cache line containing the to-be executed basic block is fetched from the L1-I cache by the IFU.  

When the control flow shifts to a new cache line, such as through a branch, the decoding process advances from the first byte of the branch target address in that new line.  That said, the taken branch leading to that line is typically not the final instruction in the cache line, nor is the branch target at the cache line's beginning.  Consequently, the bytes following a taken branch and those preceding the branch target entry point in a cache line remain undecoded during branch execution.  We note, however, that these shadow bytes are nevertheless fetched from the L1-I cache along with the rest of the cache line, though they are ultimately unused or not decoded.  These shadow bytes could contain a branch instruction that, if decoded and captured in a structure, could allow FDIP to continue running ahead of the current instruction stream when control reaches those instruction bytes in the future. We categorize these shadow branches into \textit{Head} (beginning to entry point) and \textit{Tail} (exit point to end) shadow branches.  The variability in instruction lengths within CISC architectures, makes predicting instruction boundaries difficult. Unlike RISC's (Reduced Instruction Set Computer) fixed-length instructions, such as ARM, CISC instructions can range from 1 to 15 bytes.  This complexity requires sophisticated decoding mechanisms to identify the start of valid instruction sequences accurately.

\subsection{Head and Tail Shadow Branches}
\label{sec:headtailshadow}

\begin{figure}[h]
  \centering
  \includegraphics[width=\columnwidth]{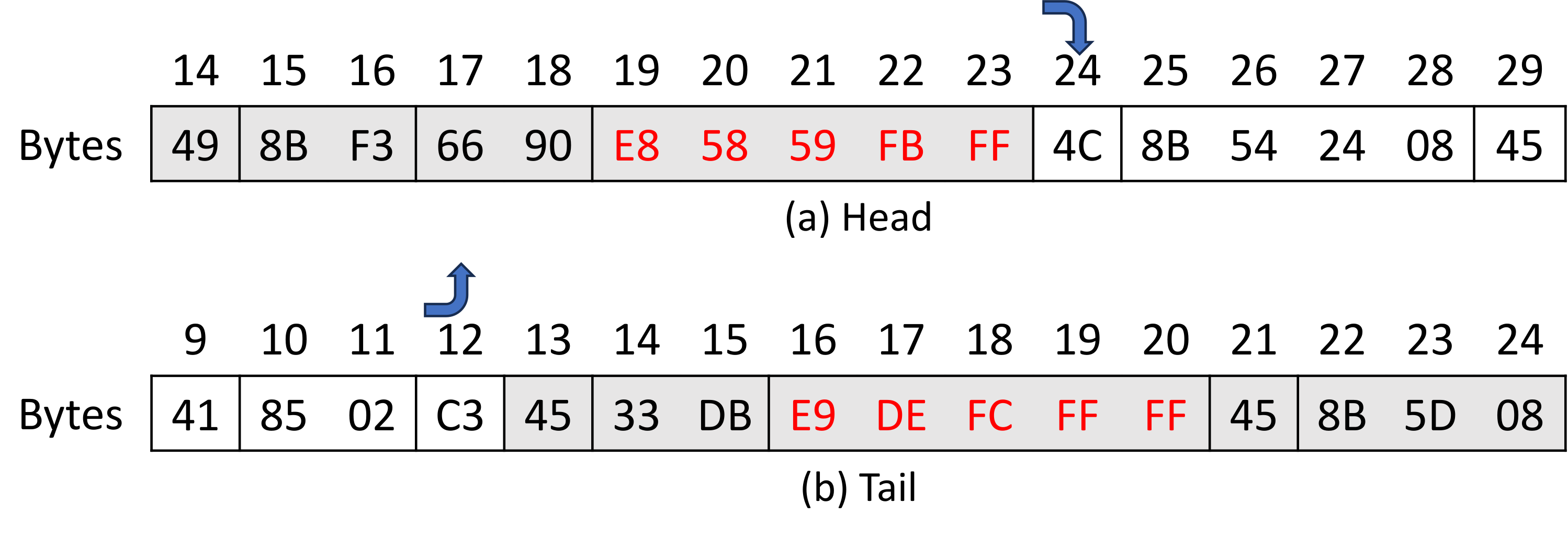}
  \caption{Two different chunks of cache lines.  Non-shaded regions represent executed instructions, while shaded regions are bytes on the cache line that are before an executed branch target or after an executed branch.  Branches in the unexecuted bytes in the lines are referred to as shadow branches, the bytes of these instructions have been colored red.}
  \label{fig:head_tail}
\end{figure}

Figure~\ref{fig:head_tail} shows the byte-by-byte contents of segments of two instruction cache lines.  The figure shows the boundaries between instructions consecutively (without border), and branch instructions are shown in red.  Figure~\ref{fig:head_tail}(a) shows a cache line segment with an entry offset of byte 24, implying that bytes 0 through 23 (shown as shaded in the figure) are fetched from the L1-I cache and fed into the decoder, only to be ignored as not part of the basic-block beginning at byte 24.  We designate branches in bytes 0 through 23 as \textit{Head} shadow branches, shown in red in the figure.

Figure~\ref{fig:head_tail}(b) presents a segment of a different cache line that exits at byte 12.  Bytes 13 to 63 are also loaded into the L1-I cache but are not sent to the decoder as the control flow directs the front-end away from these instructions before they are reached.  Any branches these bytes contain are classified as \textit{Tail} shadow branches, also shown with a red font in the figure.

\subsection{Branch Types}
\label{sec:br_types}
Branch instructions are fundamental components of a processor's Instruction Set Architecture (ISA), allowing the flow of execution to jump to different parts of a program sometimes conditionally.  There are several types of branch instructions, each serving a specific purpose.  Here is a quick rundown of the relevant types of branches which we are concerned with in this work.

\begin{description}
    \item[IndirectUnCond:] Jump to an address stored in a register or memory location.
    \item[DirectCond:] Jump to a specified address only if a certain condition is met, typically based on condition codes e.g. ``jump if zero.''
    \item[DirectUnCond:] Jump to a specified address, changing the flow of execution unconditionally.
    \item[Return:] Returns from a subroutine to the address saved by a CALL instruction. 
    \item[Call:] A form of DirectUncond Jump that saves the return address in a register or on the stack. 
\end{description}

Importantly, since we intend to opportunistically decode, and insert into the SBB, branches on the unused fragments of cache lines before and after the executed basic block, the execution time register state is not available for use in generating a target address.  Thus, only those branches where the target is determined from the PC, potentially with an encoded offset (Direct Unconditional branches and Calls) or those where it can be determined from recent Calls (Returns) are viable for our technique.

\subsection{Motivation}
\begin{figure}[h]
  \centering
  \includegraphics[width=\columnwidth]{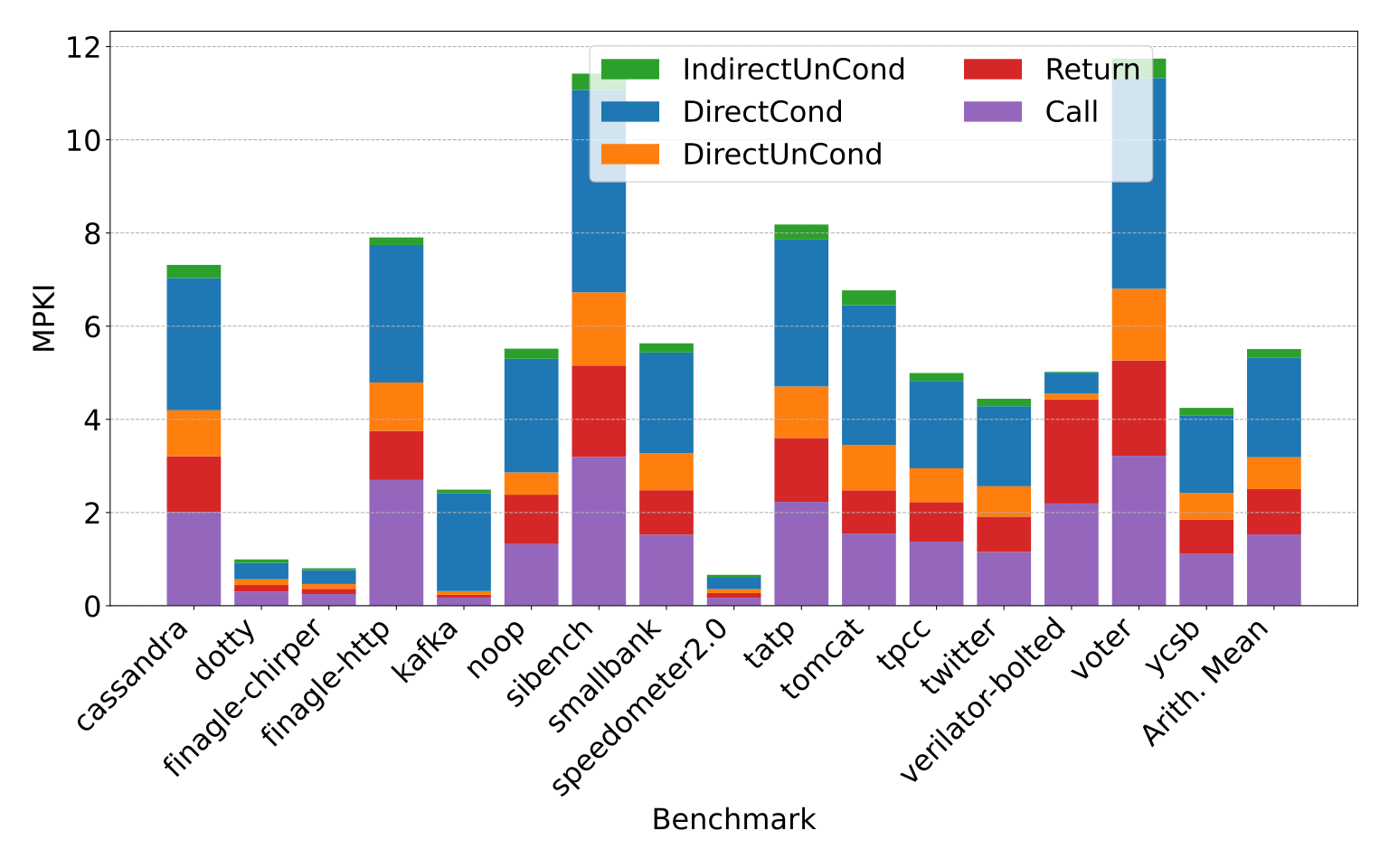}
  \caption{BTB misses by type for 8K (78KB) BTB.}
  \label{fig:BTB_Miss_all_Type}
\end{figure}

Figure~\ref{fig:BTB_Miss_all_Type} shows the breakdown of BTB misses seen in an 8K-entry BTB for each of the 16 workloads examined, by the type of branch as defined in the previous subsection.

One important, broad distinction between branches is whether they are direct or indirect.  This distinction tells us whether there is enough information between the bytes encoded in the branch instruction itself and the branch's PC to determine the target of the branch (\emph{i.e.} direct) or if more information is required, either from a register or memory (\emph{i.e.} indirect). 
 Thus, the target of a previously unseen direct (non-BTB resident) branch is generally available at decode time.  In contrast, previously unseen indirect branches require completion of the instruction in the core before the target can be known.  Looking at Figure~\ref{fig:BTB_Miss_all_Type}, we see that, at least among BTB misses, indirect branches are a vanishingly small percentage for each workload examined.  A lack of indirect branches implies that it should be possible to decode a majority of unseen branches, and we should be able to directly insert them into the BTB without requiring the actual execution of those branches to resolve first.

Referring back to Figure~\ref{fig:btb_misses_per_config}, we see that the vast majority of BTB missing branches are either Head or Tail Shadow Branches as defined in Section~\ref{sec:headtailshadow}.  Thus, putting both observations together, it should be feasible to reduce BTB misses significantly by decoding the direct, unused, Head and Tail Shadow Branches on cache lines that FDIP already brings into the core's front-end.  This is the goal of Skia.

\begin{figure*}[th!]
    \centering
    \begin{minipage}{\columnwidth}
        \centering
        \includegraphics[width=\columnwidth]{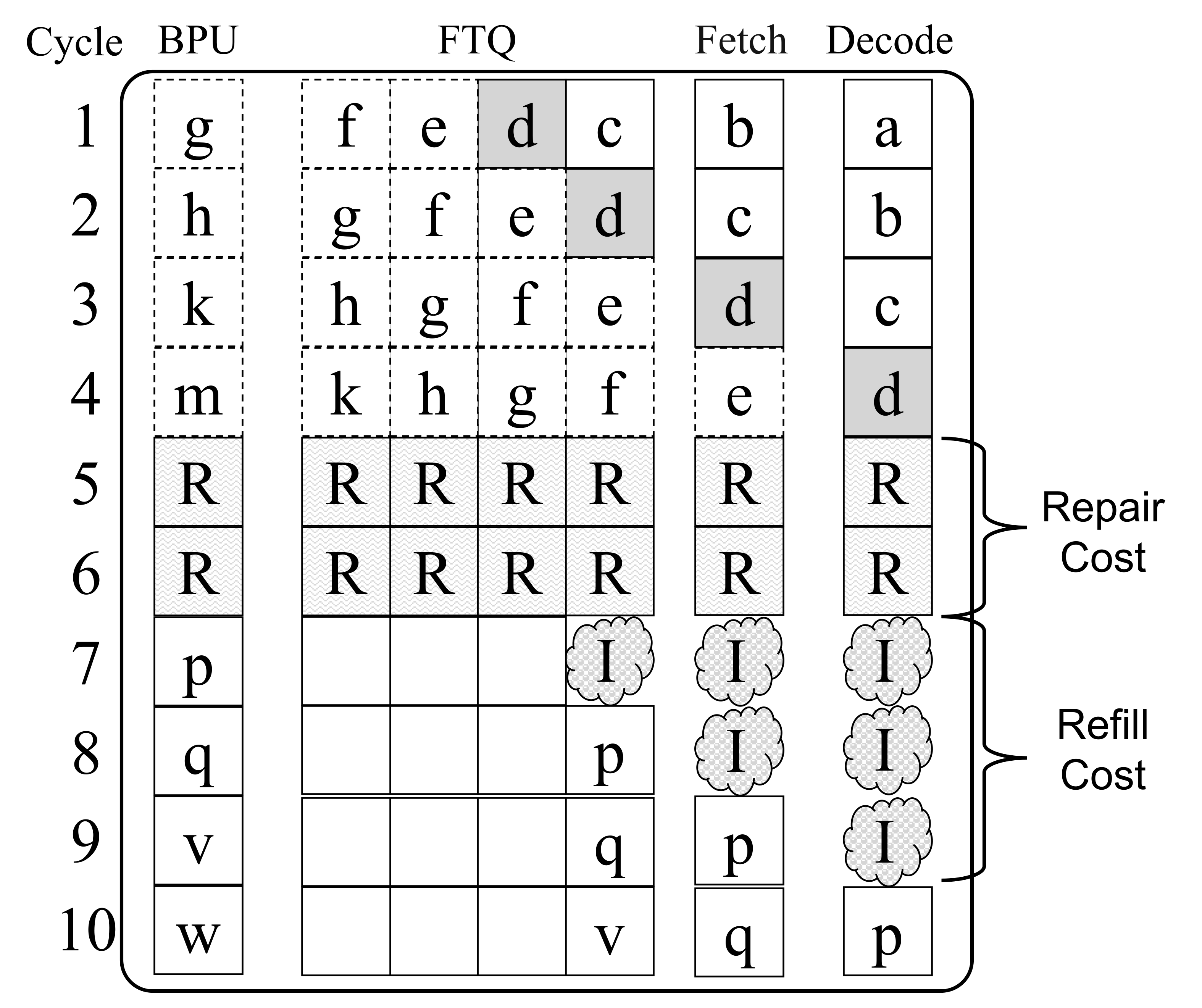}
        \subcaption{Generic Decoupled Front-End}
        \label{fig:example_i}
    \end{minipage}
    \hfill
    \begin{minipage}{\columnwidth}
        \centering
        \includegraphics[width=\columnwidth]{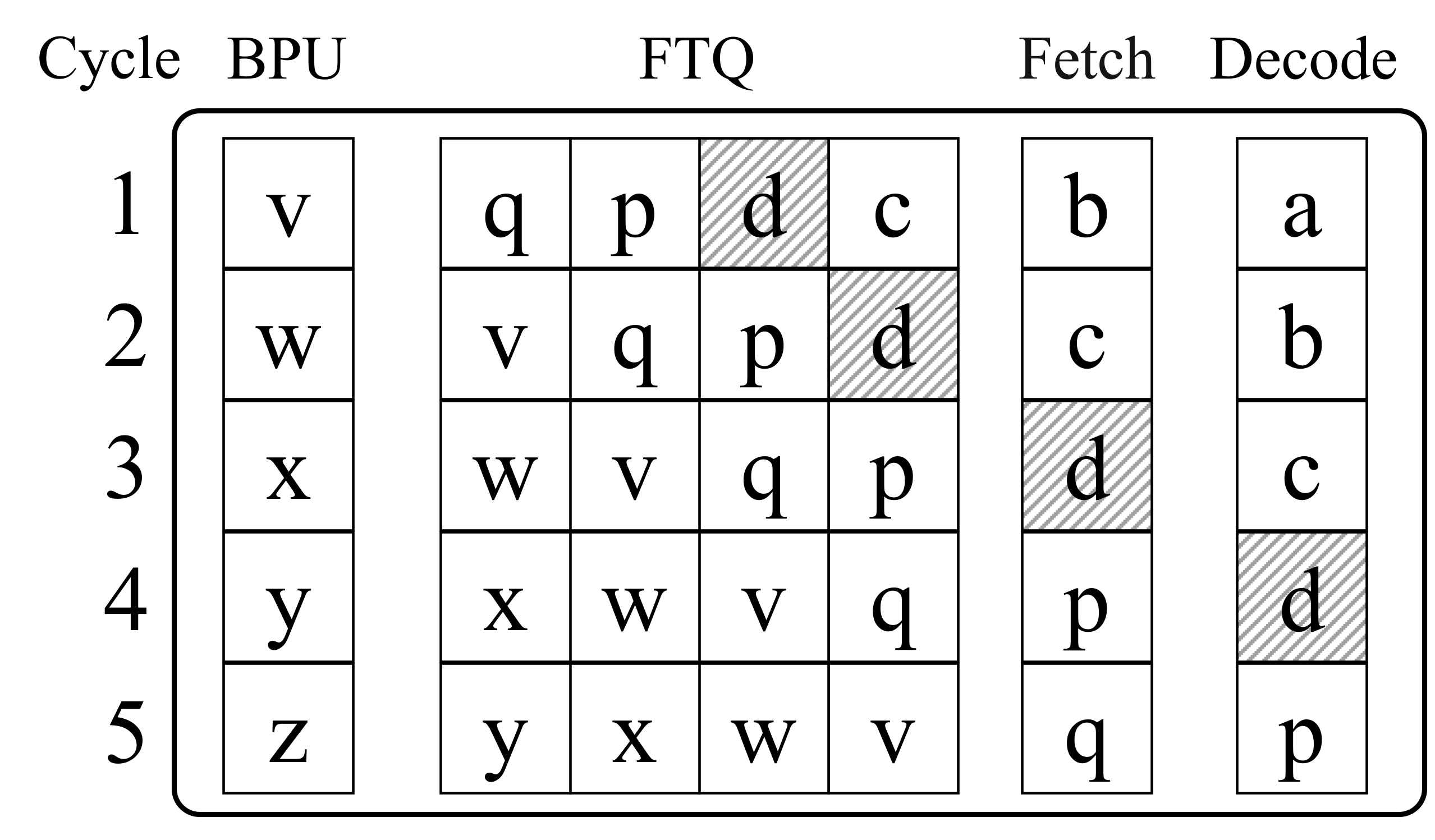}
        \subcaption{Skia Pipeline}
        \label{fig:example_ii}

        \vspace*{.1in}

        \includegraphics[width=\columnwidth]{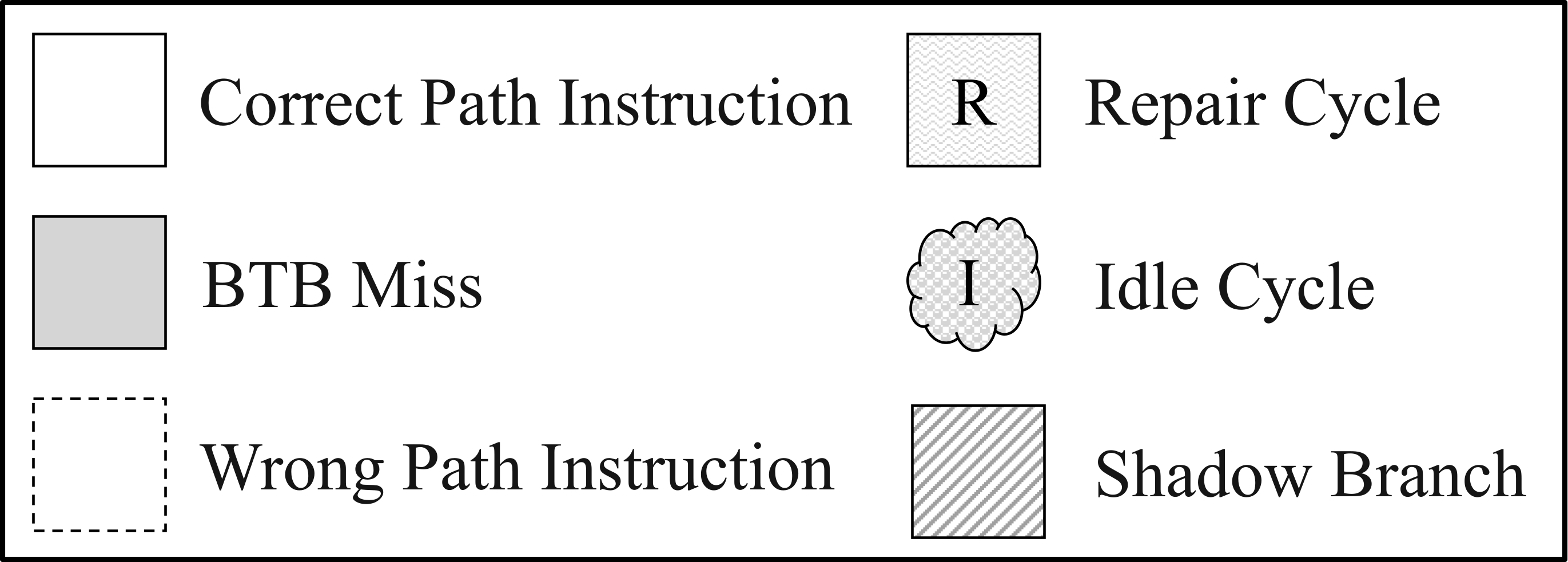}
    \end{minipage}

    \caption{Pipeline diagrams showing instructions in a standard processor front end, as well as one enhanced with Skia.}
    \label{fig:example}
\end{figure*}

\subsection{Working Example of Skia}
Figure~\ref{fig:example} presents a simplified view of the processor Front-End pipeline. In~\ref{fig:example_i} the figure letters represent instructions working through the decoupled front-end pipeline. The letter \textit{d} represents a direct unconditional branch, where the target is mispredicted due to its absence in the BTB.  With the Front-End being decoupled, the BPU continues to predict branch targets along the wrong path until it receives a corrective signal. This occurs in two scenarios: either during the Decode stage (Early re-steer) or during the Execute stage once an undetected branch is resolved. In cycle 4, the Decoder identifies the incorrect target and sends an Early re-steer signal to update the state with the correct target. Following the signal, two cycles are required to complete the repair, and by cycle 8, the correct target is made available to the FTQ. Until the correct target is delivered to the Decoder, it remains idle as the Fetch pipeline is refilled.

The impact of this misprediction is amplified if the target of branch \textit{d} results in an L1-I cache miss or if the branch resolves later in the pipeline. In such cases, the delay in correcting the prediction further stalls the Fetch and Decode stages, extending the number of idle cycles and increasing the potential for misfetched instructions to pollute the pipeline.

Figure ~\ref{fig:example_ii} shows the same pipeline with Skia. In the figure, the target of branch \textit{d} has already been identified and stored in the BPU, as shown in Figure ~\ref{fig:skia}, as the cacheline was already in the L1-I exposing the shadow branch. This preemptive identification effectively eliminates the misprediction, enabling FDIP to speculate more accurately and deeply along the correct path. This approach demonstrates the efficiency of our technique, significantly reducing idle cycles in the Decode stage and minimizing the number of instructions that pollute the cache by following incorrect paths.

\section{Skia Design}
\label{sec:skia}
In the context of commercial and data center workloads, the issue of instruction stream resteers resulting from BTB misses on infrequently encountered, or 'cold', branches is a significant challenge that needs to be addressed.  Previous solutions attempt to mitigate this issue by prefetching into the BTB, which necessitates substantial hardware modifications or software profiling but does not adequately address a significant portion of cold BTB misses.  As discussed previously, we observe that the branches that miss in the BTB are consistently already in the L1-I cache, having been fetched to execute other basic blocks containing instructions in the same cache lines.  However, the missed branches remain un-decoded because they are overshadowed by an executed branch that leaves the cache line or by being before the target of a branch into the cache line.  Surprisingly, the bytes encoding these shadow branches are often already pulled into the processor front-end by FDIP but they are discarded as being off the current executed path.

This section describes our approach to identifying and decoding shadow branch instructions in a variable-length ISA.  As discussed in Section~\ref{sec:background}, each cache line can contain either unused Head or Tail bytes, requiring different approaches for Head and Tail decoding.  The following subsections discuss our approach to decoding those Head and Tail Shadow Branches and the microarchitectural modifications for storing them until their use. Since Skia only decodes branches on the cache line of already executing code, there is no implied privilege violation.  Further, since Skia only decodes direct branches and returns it would be difficult for an attacker to manipulate the target address of the shadow branch for malicious purposes.

\subsection{Discovering Shadow Branches}
Identifying and decoding shadow branches varies depending on whether they are Head or Tail Shadow Branches, with Head Shadow Branches being significantly more challenging to identify. 
 Here, we will discuss the identification and decoding processes for each type of shadow branch in detail.

By opportunistically decoding bytes from the beginning of the cache line up to the entry offset (the target of the jump instruction) within that cache line, we can broadly identify Head Shadow Branches.  However, Tail branches are identified by decoding bytes starting from a taken branch and continuing to the end of the cache line. 

\begin{figure}[h]
  \centering
  \includegraphics[width=\columnwidth]{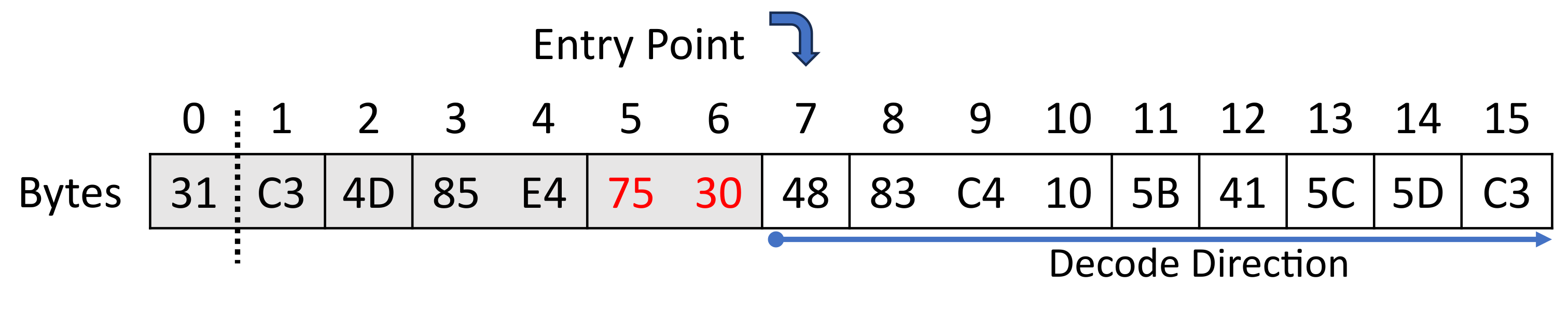}
  \caption{When commencing the decoding process from the 0-byte, it reveals a valid 2-byte instruction, specifically \textit{xor ebx, eax}.  Conversely, initiating decoding from byte 1 also results in a valid instruction, namely a \textit{ret}.}
  \label{fig:head}
\end{figure}

\subsection{Head Shadow Branch}
\label{sec:head_shadow_branch}
Identifying Head Shadow Branches poses significant challenges in CISC ISAs.  With variable-length instruction encoding, identifying instructions looking backward from the branch target may yield more than one possible set of instructions.  Figure~\ref{fig:head} illustrates the problem.  The figure shows the entry point at byte 7 in the line and the shadow region covering bytes 0-6 in the line.  Critically, as the figure shows, this case has two possible valid decodings of instructions in the shadow region.  This is possible because we do not know if byte 0 in the cache line is the beginning or somewhere in the middle of a variable-length instruction.  In this particular case, two possible sets of instructions could be decoded out of bytes 0-6; the first starts with an instruction at byte 0 of the cache line, and the second starts with an instruction at byte 1 of the cache line.  Of course, only one of these decodings is actually "true" (here the \emph{ret} is a bogus branch).  Interestingly, in this particular case both "paths" (\emph{i.e.} starting decode from either byte 0 or 1), converge after the first instruction and the shadow branch (highlighted in red) will be correctly decoded.  We call when different decode paths converge to one path, "merging path" and discuss it further below.

For Head decoding, we specifically target cache lines related to the start of the FTQ entry.  This targeting is strategic because the FTQ entry's beginning corresponds to the target of a branch, as each entry represents a continuous set of instructions.  The observation discussed above that instructions may not start at the beginning of the cache line prompts us to target these specific cache lines for head shadow branch decoding. 

\begin{figure}[h]
  \centering
  \includegraphics[width=\columnwidth]{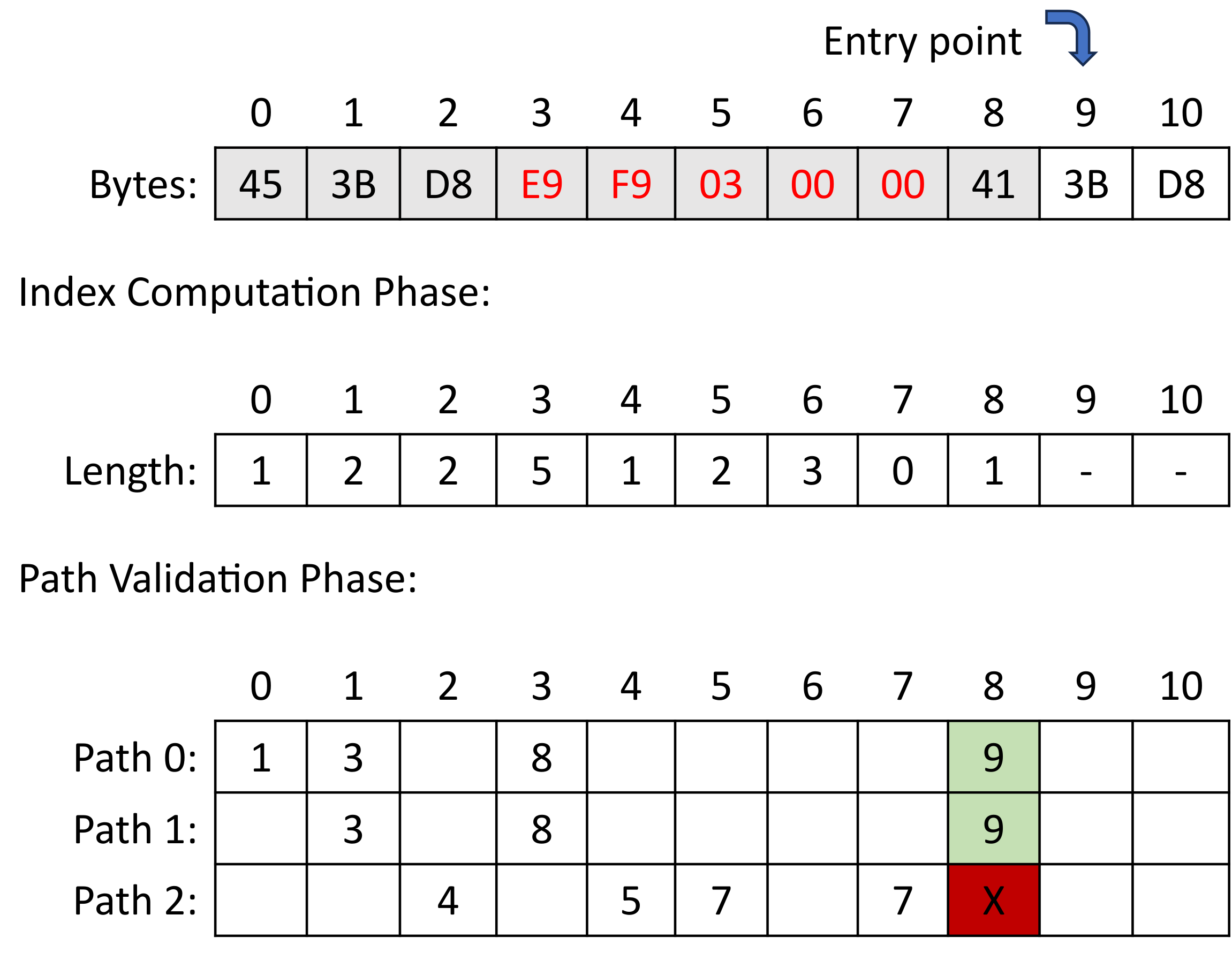}
  \caption{An example of how the Index Computation and Path Validation phases work as described below.}
  \label{fig:phases}
\end{figure}

The Shadow Branch Decoding process initiates upon completing the prefetch request and confirming that the cache line is present in the L1-I cache\footnote{We emphasize that shadow branch decoding is far off the critical path of the front-end. It is done in parallel with the regular FDIP-to-Decode path, as the branches decoded are typically not used for some time after the initial executed path decode of the line. Thus, the process can take multiple cycles.}.  This process consists of two main phases: Index Computation and Path Validation.  The first phase focuses on annotating the instruction boundaries, while the second phase is concerned with identifying direct unconditional branches and returns.  These stages are structured into distinct stages to optimize the Head Decoding process.  These phases are illustrated in Figure~\ref{fig:phases}.

The first phase, identifies the instruction boundaries within the cache line to determine the beginning of the target segment for shadow branch decoding.  This stage involves computations and analytical processes to pinpoint the potential byte offset within the cache line where shadow branch decoding should start from, as elaborated in the following section.  Once the Index Computation stage identifies the start offset, it initiates a Path Validation phase, focusing on decoding bytes from this start index up to the branch's target.  The target/end-byte offset is identified from the FTQ information of the cache line. 

\subsubsection{Index Computation}
\label{sec:index_computation}
The Index Computation phase determines the byte length of potential instructions within a byte stream.  This process begins by sequentially feeding bytes from position 0 to the decoder until it detects a complete instruction.  We record the length of this instruction in a vector called \textit{Length}.  Once the length is recorded, the index is incremented, and the process repeats, continuing until the entry offset is reached.  This iterative approach ensures that every potential instruction starting from each byte position is accurately measured and recorded.  

Referring to Figure~\ref{fig:phases}, the decoding process reveals that byte 0, represented by the hexadecimal 45, is identified as a single-byte instruction.  In a similar vein, the decoder requires a sequence of 5 bytes to successfully decode an instruction commencing at byte 3, indicating the start of a potential instruction spanning 5 bytes (E9 F9 03 00 00).  Progressing to the subsequent byte (F9), the decoder is capable of generating an instruction consisting of just one byte.  In this scenario, there is an overlap, making it impossible for both decodings to be correct; therefore, we need to validate them.  Finally, the presence of a zero in the figure denotes the inability to decode a valid instruction from that specific byte, for instance, at byte 7.

\subsubsection{Path Validation}
\label{sec:path_validation}
In the Path Validation phase, the information derived from the Index Computation phase is employed to identify shadow branches within all potential instruction sequences.  The process begins by constructing a path starting with the value at index 0 of the Length vector.  Subsequently, the index is incremented by this retrieved length value, and the newly acquired length is added to the path.  This iterative procedure continues until the path aligns with the entry point of the line.  If the path aligns, indicating the accuracy of the Index Computation phase, we check for the presence of any supported branches (jmp, call, return) and insert them into the corresponding SBB structure.

For example, in Figure~\ref{fig:phases}, we start with \textit{path} = \textit{Length}[0] = 1, then \textit{path} += \textit{Length}[\textit{path}] = 3, continuing this process until \textit{path} is equal to the entry offset if possible. Path 0 and Path 1 lead to a correct path, but Path 2 does not as \textit{path} = \textit{Length}[2] = 2, then \textit{path} += \textit{Length}[\textit{path}] = 4, leading to index 7 which is not a valid instruction.

It's evident that Path 0 and Path 1 share the same path from a specific point onward, this creates a "merging path".  Based on this observation, we introduce optimizations to enhance the accuracy of decoded shadow branches:
\begin{description}
    \item[Valid Encodings:] During path generation, if a maximum of six valid paths is reached based on empirical selection criteria, the associated cache line is discarded. This method ensures thorough exploration of potential instruction chains while effectively managing computational resources. \emph{i.e.} We have 3 valid paths (path: 0, 1, 3) in Figure~\ref{fig:phases}.
    \item[Valid Index:] Observing that numerous valid paths converge into a merge path, we examined which path would yield the best performance. Empirical selection revealed that consistently using the First Index provides better results compared to using the Zero Index or the Merge Index. We define the First Index as the index where the first valid path is found, the Zero Index denotes the point where, upon finding a valid path, byte decoding begins starting from index zero, and the Merge Index as the most common recent index among all valid paths.  In Figure~\ref{fig:phases} for example, the First Valid Index is equal to Zero Index = 0 (Byte 45) and the Merge Index = 3 (byte E9).

\end{description}

These optimizations contribute to the accurate decoding of shadow branches and improve the overall efficiency of the Path Validation phase. On average, Skia introduces 0.0002\% additional bogus branches into the FDIP stream relative to the total entries added to the SBB. On a related note, since the branches that Skia decodes and places into the SBB are only direct branches and returns, it would be difficult for an attacker to manipulate Skia to achieve information leakage or other forms of attack.

\begin{figure}[h]
  \centering
  \includegraphics[width=\columnwidth]{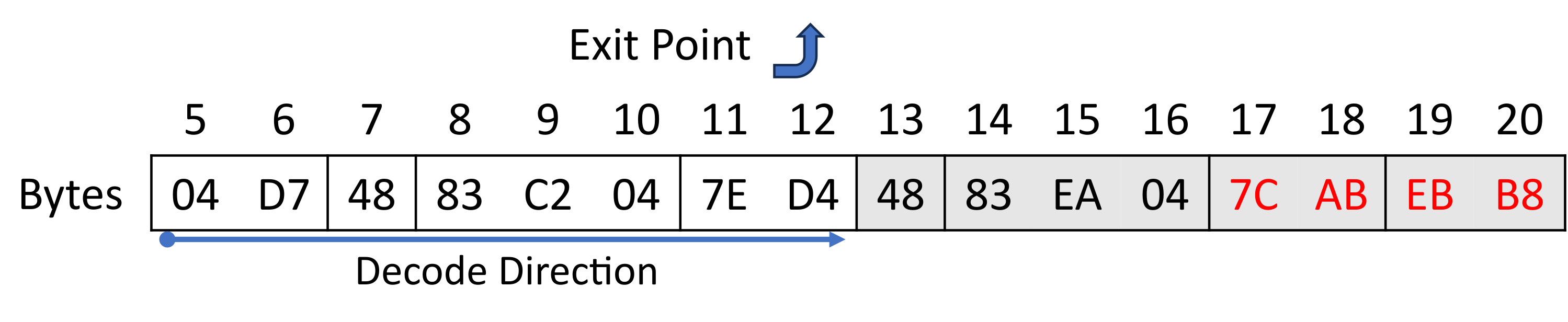}
  \caption{Tail Shadow Branch Decoding: Once we decode all the bytes until the end of the cache line the highlighted branch in red is going to be inserted into SBB.}
  \label{fig:tail}
\end{figure}

\subsection{Tail Shadow Branch}
For Tail shadow branch decoding, we specifically target cache lines that mark the end of the FTQ entry. The end of the FTQ entry is denoted by a branch instruction that redirects the control flow away from that particular cache line. This provides the opportunity to decode the remaining shadow bytes. These bytes typically remain undecoded as the control flow exits the cache line, making them prime candidates for efficient shadow branch decoding.

Figure~\ref{fig:tail} illustrates Tail Shadow Branch Decoding.  As the figure shows, because the branch ending the FTQ entry is known, the start byte of the first instruction in the shadow region is also known. Thus there is only one possible set of instructions in the shadow region making the decode process far more straight forward.

\subsection{Head v/s Tail Shadow Branch Decoding}
Discovering head shadow branches involves computational steps that can occasionally yield incorrect results, potentially leading to an incorrect start point for decoding. This can result in the decoding of instructions that do not actually exist in the program's flow. Furthermore, these incorrect instructions might contain nonexistent, or "bogus", branches that could adversely affect the BPU, leading to inaccuracies in branch prediction and thus, performance degradation.

When it comes to discovering tail shadow branches, the challenge of determining a starting point for decoding is less pronounced. This is because we already know the start and end points of the taken branch instruction. Therefore, we can begin decoding from the end of the branch instruction until the end of the cache line without the ambiguity that arises with head shadow branches. This clarity in determining the decoding start point simplifies the process for tail shadow branches compared to head shadow branches, where the variability in instruction lengths and the presence of prefixes make identifying the beginning of a valid instruction sequence more complex.

The design and implementation of discovering head and tail shadow branches are orthogonal, meaning they are independent and can be utilized separately or in combination based on power, area budget and performance requirements. This flexibility allows architects to choose between focusing on discovering head shadow branches, tail shadow branches, or both, depending on the specific needs of their system. 
In Section \ref{sec:evaluation}, the performance impact of discovering just head shadow branches, just tail shadow branches, and both combined is described and discussed in detail.

\section{Design Implementation}
\label{sec:design_implementation}
Figure \ref{fig:skia} illustrates the integration of Skia's components into the BPU, including the Shadow Branch Decoder (SBD) and the Shadow Branch Buffer (SBB). When the SBD identifies a supported branch instruction, it inserts it into the corresponding SBB. During a BTB lookup, the SBB is accessed concurrently. If a BTB miss occurs, the SBB will supply a target if one is available. Each of these components is described in detail below.

\subsection{Shadow Branch Decoder (SBD)}
\begin{figure}[h]
  \centering
  \includegraphics[width=\columnwidth]{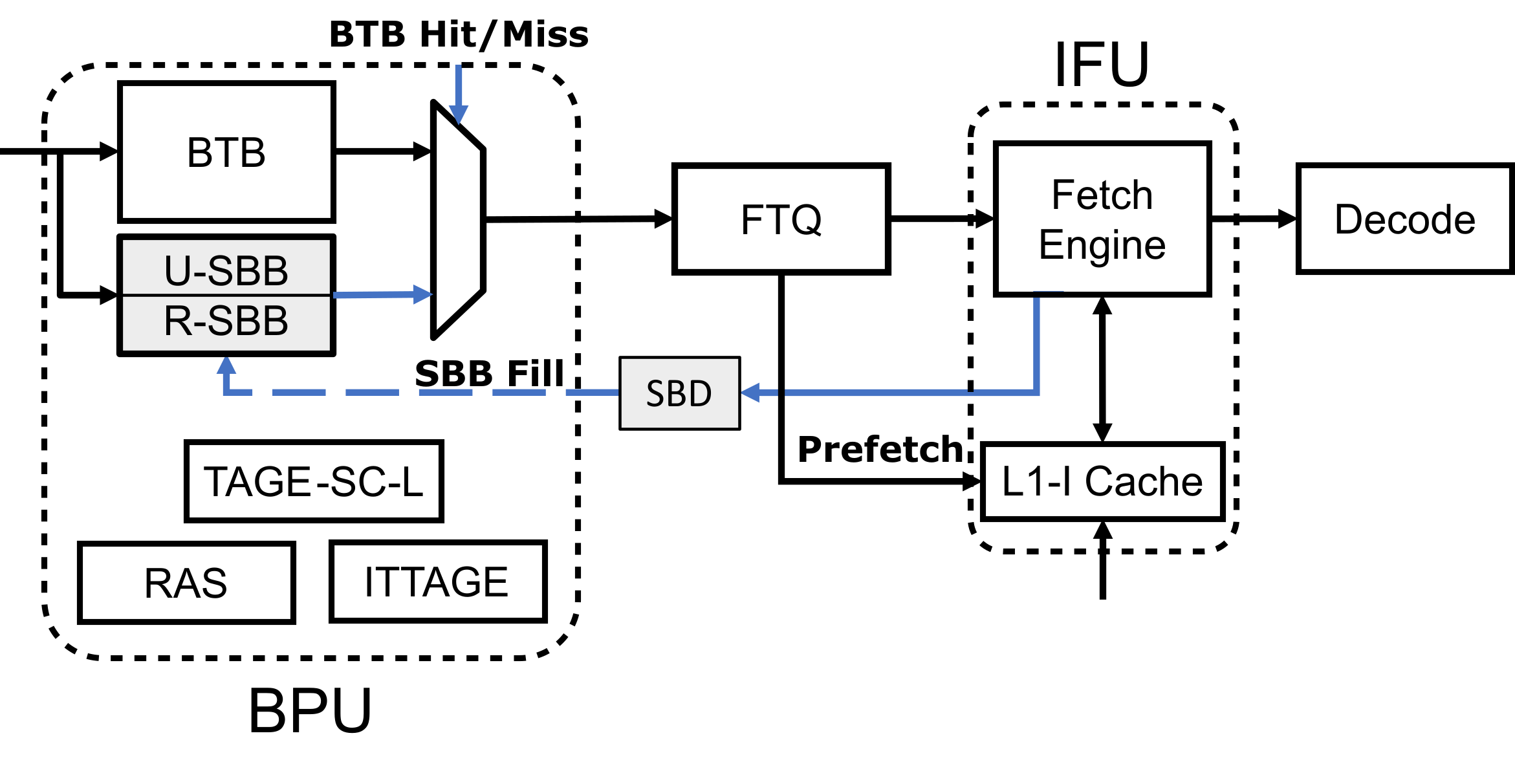}
  \caption{Proposed Skia Design.}
  \label{fig:skia}
\end{figure}

The Shadow Branch Decoder (SBD) is a highly simplified decoder focused solely on identifying instruction boundaries and decoding supported branch instructions.
Upon fetching a new cache line, the SBD scans the bytes using the algorithms discussed in Section~\ref{sec:skia} above for decoding head and tail shadow branches. When SBD identifies a branch, it is inserted into the SBB as shown in Figure~\ref{fig:skia}.

\subsection{Shadow Branch Buffer (SBB)}
One might assume that once the SBD identifies shadow branches, they should be inserted directly into the BTB. However, the BTB is a critical path component of the front-end, which we want to avoid taking bandwidth away from and prevent inserting possibly incorrect branches into, causing pollution.

We propose a novel parallel structure to the BTB that delivers significant performance enhancements despite its simplicity and compact size. Figure~\ref{fig:skia} illustrates the division of the SBB into two distinct buffers: the DirectUncond SBB (U-SBB) and the Return SBB (R-SBB). The U-SBB exclusively stores Direct Unconditional branches, whereas the R-SBB is dedicated to Return instructions. This approach allocates specific roles to the U-SBB and R-SBB, optimizing both area utilization and performance.

\begin{figure}[h]
    \centering
    \includegraphics[width=\columnwidth]{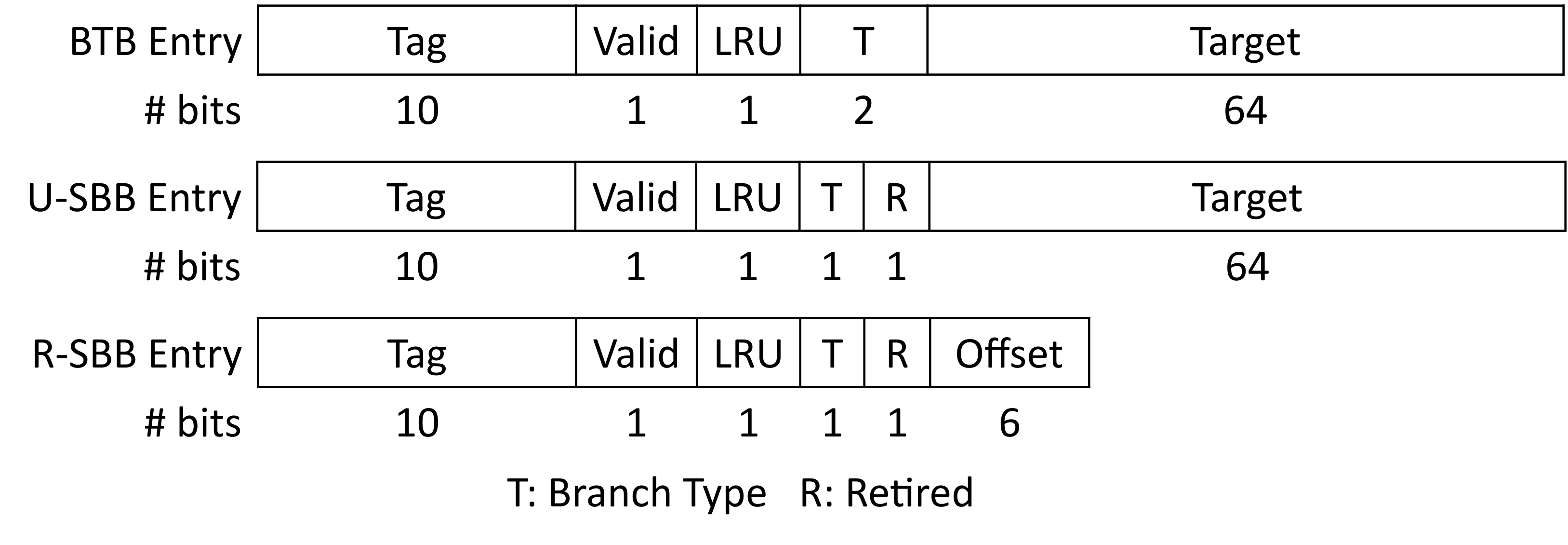}
    \caption{An entry of each structure and the required fields.}
    \label{fig:entries}
\end{figure}

Figure~\ref{fig:entries} depicts the structure of each entry.  An entry consists of 10 bits designated for the tag, 1 bit to indicate validity, and 1 bit for the Least Recently Used (LRU) status per way.  BTB entries allocate 2 bits for identifying the branch type and 64 bits for the branch target address.  U-SBB entries utilize 1 bit to mark retired instructions and 64 bits for the target address.  R-SBB entries use 6 bits for the offset and 1 bit to denote retirement.  In total, an entry in the R-SBB requires only 20 bits compared to 78 bits for an entry in the U-SBB. This efficiency allows the SBB structure to retain more overall entries.

\subsection{Replacement policy}
In the implementation of the SBB structures, the Least Recently Used (LRU) replacement policy is utilized. Additionally, when a branch target provided by the SBB is committed, the "Retired" bit is set in the corresponding SBB entry. This ensures that "bogus" branches are evicted first, allowing the "useful" branches to remain longer.

\section{Simulation Methodology}
\label{sec:simulation_methodology}
The section outlines the simulation framework, the large instruction footprint workloads and different configurations to evaluate Skia, our shadow branch decoding technique.

\subsection{Baseline Simulation Model}
Our baseline CPU setup mimics the Golden Cove ~\cite{wikichip} (commercially known as Alder Lake) CPU core microarchitecture using the gem5 simulator~\cite{gem5}, as detailed in Table \ref{tab:processor_configurations}.
The study employs simulating workloads through an out-of-order, execution-driven CPU model (O3CPU) within Full system simulation, which emulates a complete operating system (Ubuntu) and runs multi-threaded JAVA applications. The O3CPU has been extended to model branch-misprediction-based wrong path execution. Initially, the workloads undergo a warm-up phase of around 10 million instructions, during which the caches, branch predictor, and other structures are primed. Following this warm-up phase, the simulation transitions to detail mode (O3CPU) and continues for an additional 100 million instructions.  Also, we used the cacti~\cite{cacti} tool to approximate the latency as the BTB scales.

\subsection{Core Front-End Modeling}
One notable contribution and unique aspect of this work is our faithful modeling of an exceptionally aggressive processor front-end. We achieve this by extending gem5's O3CPU model to incorporate FDIP, enabling support for a decoupled front-end. 

Given that FDIP's performance directly depends on the branch predictor's accuracy, we enhanced gem5's BPU by integrating TAGE-SC-L~\cite{tage_scl} with an ITTAGE indirect predictor~\cite{ittage} and employing an 8K-entry BTB. We also integrate support for the BPU indirect predictor and the BTB to queue predicted cache lines into the FTQ. The FTQ is capable of directly issuing prefetches into the L1-I cache.

In the event of control flow resteers, the FTQ is flushed before resuming fetching from the correct path. Since gem5 operates in an execution-driven manner, it accurately models the effects of such wrong-path resteers. 

In our baseline setup, we employ a 24-entry FTQ, with each entry corresponding to a basic block. This configuration strikes a balance by providing enough depth to tolerate miss latency while avoiding excessive front-end aggressiveness that could lead to adverse effects. 

The commercial processor vendors have been using FDIP-based front-end designs for over a decade, as evidenced by recently disclosed commercial CPU designs~\cite{AMD_Zen2, 8986666, 9138988, rupley2018samsung}. Ishii et al.~\cite{ishii2020rebase}
raised similar concerns regarding necessity of using FDIP for modern front-end
research. Therefore we used gem5 with FDIP model using in PDIP~\cite{godala2024pdip}
with further enhancements are the baseline in our work.

\begin{table}[h]
\centering
\begin{tabular}{ll}
\multicolumn{1}{c}{Field / Model}         & \multicolumn{1}{c}{Alder Lake like}     \\ \hline
\multicolumn{1}{|l|}{ISA}                 & \multicolumn{1}{l|}{X86}                \\ \hline
\multicolumn{1}{|l|}{Private L1-I Cache}  & \multicolumn{1}{l|}{32KB (8-way, 64B)}  \\ \hline
\multicolumn{1}{|l|}{Private L1-D Cache}  & \multicolumn{1}{l|}{64KB (16-way, 64B)} \\ \hline
\multicolumn{1}{|l|}{Private L2 Cache}    & \multicolumn{1}{l|}{1MB (16-way, 64B)}  \\ \hline
\multicolumn{1}{|l|}{Shared L3 Cache}     & \multicolumn{1}{l|}{2MB (16-way, 64B)}  \\ \hline
\multicolumn{1}{|l|}{Branch Predictor}    & \multicolumn{1}{l|}{\begin{tabular}[c]{@{}l@{}}TAGE-SC-L~\cite{tage_scl} (64KB)\\ ITTAGE ~\cite{ittage}(64KB)\end{tabular}} \\ \hline
\multicolumn{1}{|l|}{BTB Size}            & \multicolumn{1}{l|}{8K-entry/78KB (4-way)} \\ \hline
\multicolumn{1}{|l|}{U-SBB Size}          & \multicolumn{1}{l|}{7.3125KB (4-way)} \\ \hline
\multicolumn{1}{|l|}{R-SBB Size}          & \multicolumn{1}{l|}{4.9375KB (4-way)} \\ \hline
\multicolumn{1}{|l|}{FTQ}                 & \multicolumn{1}{l|}{24 Entries}         \\ \hline
\multicolumn{1}{|l|}{Decode / Retire}     & \multicolumn{1}{l|}{12 Wide}            \\ \hline
\multicolumn{1}{|l|}{ROB Entries}         & \multicolumn{1}{l|}{512}                \\ \hline
\multicolumn{1}{|l|}{Issue / Load / Store Queue} & \multicolumn{1}{l|}{194 / 144 / 112}                                                     \\ \hline
\multicolumn{1}{|l|}{Int / Vec. Registers} & \multicolumn{1}{l|}{448 / 400}          \\ \hline
\end{tabular}
\caption{Processor configurations}
\label{tab:processor_configurations}
\end{table}

\subsection{Benchmarks}
\begin{figure}[h]
  \centering
  \includegraphics[width=\columnwidth]{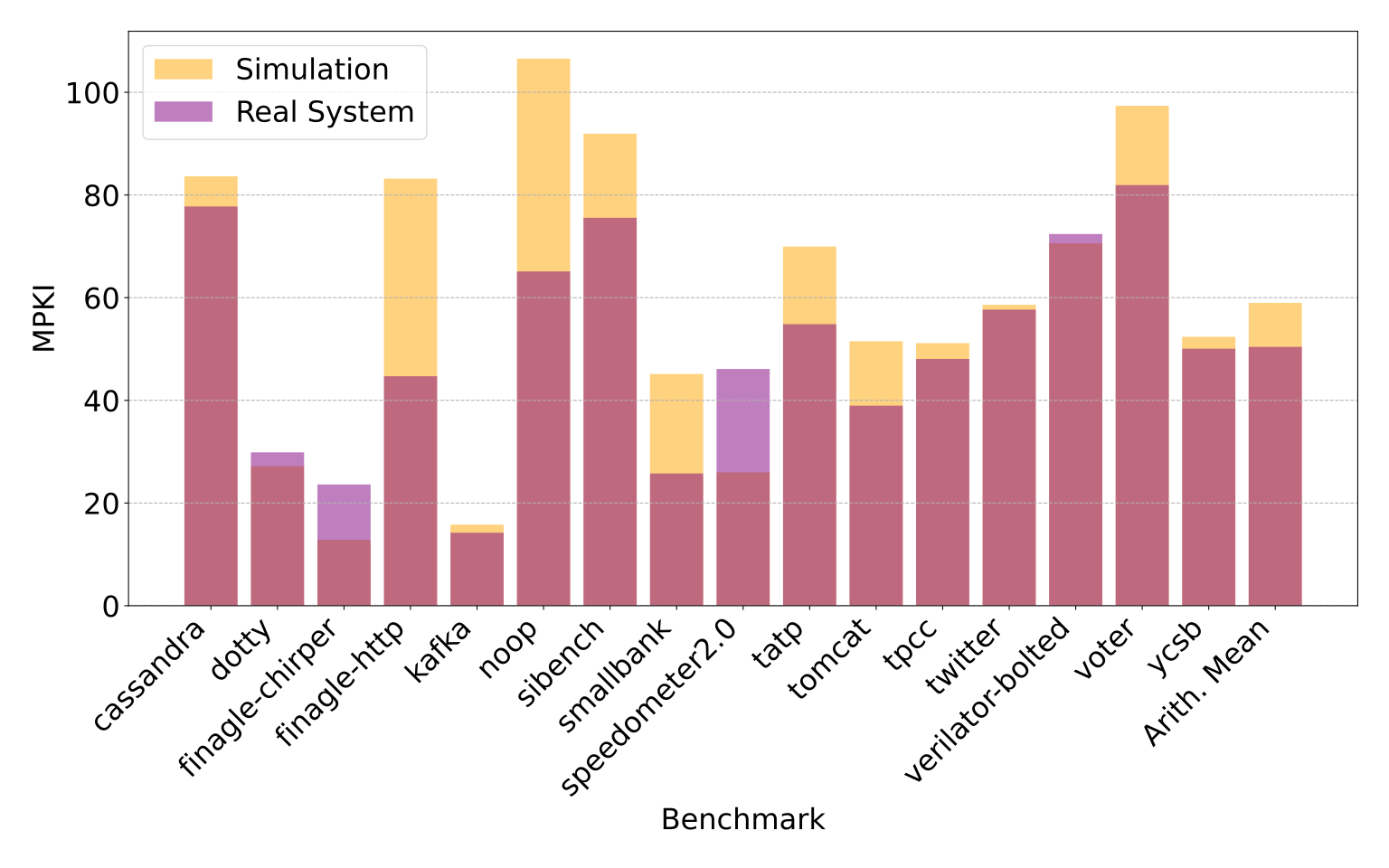}
  \caption{Comparison of L1-I MPKI between real system and gem5 simulation.}
  \label{fig:l1i_mpki}
\end{figure}

We evaluated our approach using 16 widely used client-side and server-side multi-threaded workloads with substantial code footprints, thus stressing the CPU front-end. These workloads were selected from various benchmark suites, as listed in Table \ref{tab:benchmarks} \cite{dacapo_bench,java_renaissance,oltp_bench,chipyard, browser_bench}.  Benchmarks with an L1-I MPKI of over 10 are used in this work as shown in Figure ~\ref{fig:l1i_mpki}.
We used Intel’s VTune Profiler~\cite{vtune} on a modern Alder Lake Intel processor to run benchmarks on a real machine to determine when a given workload reaches its steady state, creating a gem5 checkpoint at this point for our simulations. Figure~\ref{fig:l1i_mpki} compares the MPKI at the L1-I cache level between a real system and a gem5 simulation across various benchmarks. Overall, the simulation exhibits very similar MPKIs relative the real system, with the total difference across benchmarks being below 18\%, indicating that the simulation provides a relatively close approximation of real system behavior. This small difference suggests that the simulation results are reasonably aligned with the real hardware, despite some discrepancies. Notably, the simulation results represent check-pointed runs, while the real system results reflect longer simulation in similar region of interests of each benchmark.

BOLT~\cite{bolt}, is a relatively recent software technique where the binary is instrumented and then profiled and this profiling data is used to improve the instruction cache and BTB behavior.  By its nature it can only be applied to pre-compiled binaries, thus of the applications we examined it was only applied to Verilator (hence all results to this point are shown as "verilator-bolted").  For completeness we also examined our Skia technique on a non-bolted version of Verilator.

\begin{table}[h]
\centering
\begin{tabular}{|ll}
\multicolumn{1}{c}{Benchmark Suite} & \multicolumn{1}{c}{Benchmarks}     \\ \hline
\multicolumn{1}{|l|}{DaCapo~\cite{dacapo_bench}} & \multicolumn{1}{l|}{cassandra \cite{cassandra}, kafka \cite{kafka}, tomcat \cite{tomcat}} \\ \hline
\multicolumn{1}{|l|}{Renaissance~\cite{java_renaissance}} & \multicolumn{1}{l|}{\begin{tabular}[c]{@{}l@{}}finagle-chirper, finagle-http~\cite{twitter_finagle}, \\dotty~\cite{dotty}\end{tabular}} \\ \hline
\multicolumn{1}{|l|}{\begin{tabular}[c]{@{}l@{}}OLTB Bench~\cite{oltp_bench} \\(PostgreSQL~\cite{postgres})\end{tabular}} & \multicolumn{1}{l|}{\begin{tabular}[c]{@{}l@{}}tpcc~\cite{tpcc}, ycsb~\cite{ycsb}, twitter, voter, \\smallbank, tatp, sibench, noop\end{tabular}} \\ \hline
\multicolumn{1}{|l|}{Chipyard~\cite{chipyard}} & \multicolumn{1}{l|}{verilator\cite{verilator}} \\ \hline
\multicolumn{1}{|l|}{Browser Bench~\cite{browser_bench}} & \multicolumn{1}{l|}{speedometer2.0~\cite{speedometer2.0}} \\ \hline
\end{tabular}
\caption{Benchmarks used to evaluate Skia.}
\label{tab:benchmarks}
\end{table}

\begin{figure*}[th!]
\centering
    \includegraphics[width=2\columnwidth]{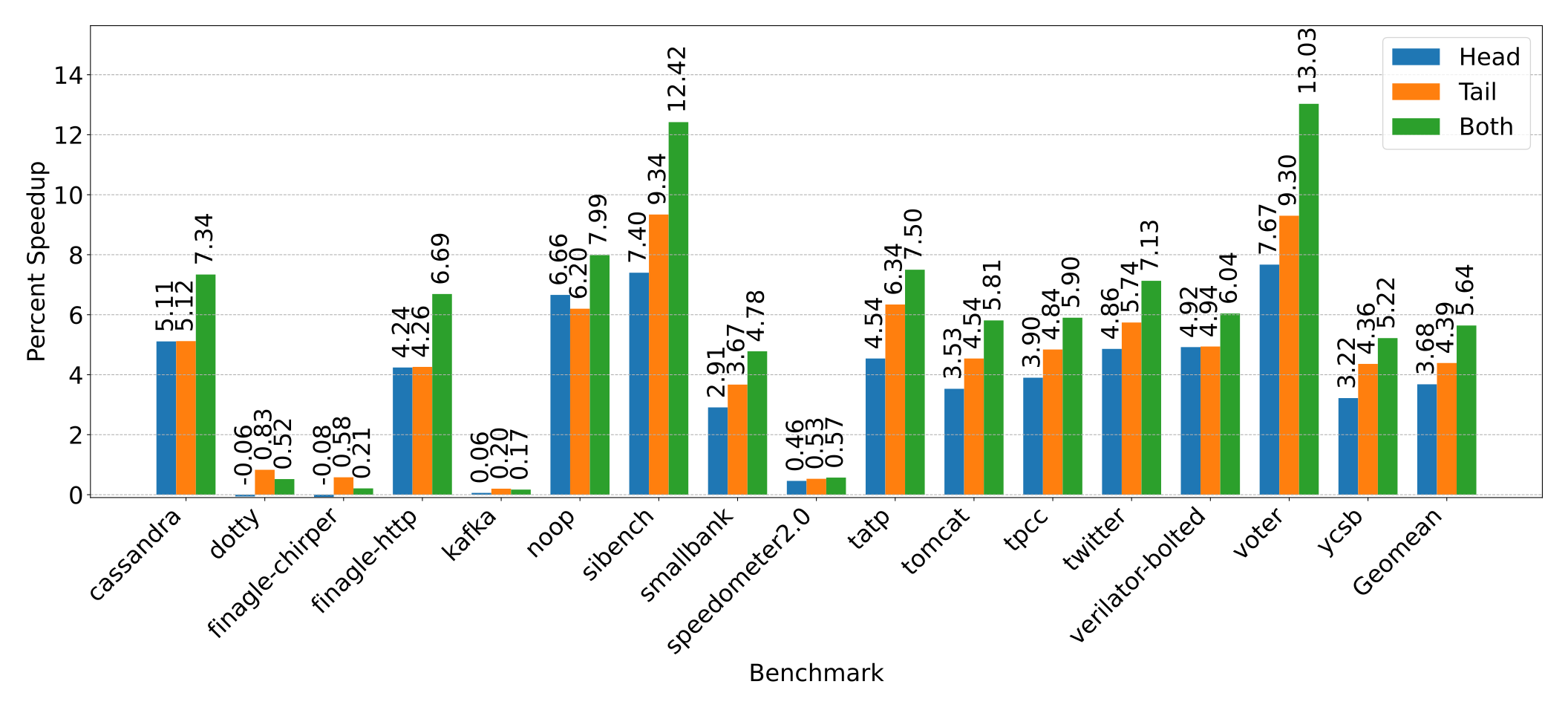}
    \caption{IPC performance gain across different benchmarks over 8K-entry (78KB) BTB.}
    \label{fig:ipc_gain}
\end{figure*}

\subsection{OS and IO bottlenecks: Full System}
In Full System simulation the kernel being simulated is responsible for software context switches which adds potential noise when multi-threaded workloads are used.  In addition to the kernel scheduler noise, IO interrupts also trigger context switch which is another source of noise.  To address these concerns we have similar approach proposed in PDIP~\cite{godala2024pdip}.  We have ensured that divergence between different configurations is within 0.2\%.

\section{Evaluation}
\label{sec:evaluation}
This section discusses the results of our evaluation of Skia using the described framework and methodology. We emphasize Skia's impact on system performance and its effects on L1-I and BTB.

\subsection{Performance Analysis}
Figure \ref{fig:ipc_gain} depicts the relative IPC gains observed across our benchmark suite under different configurations: head-only, tail-only, and combined (head and tail) opportunistic shadow decoding. Overall, shadow decoding demonstrates a significant geomean speedup of 5.64\% compared to the baseline performance of FDIP. 

Interestingly, head shadow decoding yields a 3.68\% geomean speedup, while tail shadow decoding alone achieves a respectable, higher improvement in performance of 4.39\%.  Given the complexity of implementing head shadow decoding, due to the non-determinism of the paths, designers may choose to only implement tail shadow decoding and achieve most of the performance benefit. 

\subsubsection{Performance analysis with respect to BTB Misses}
Benchmarks \texttt{finagle-chirper}, \texttt{kafka}, and \texttt{speedometer2.0} show a lower IPC gain relative to the others.  We note that these benchmarks also have lower total BTB misses, as illustrated in Figure \ref{fig:BTB_Miss_all_Type}. The lack of shadow branches considerably narrows the potential and impact of the opportunistic decoding technique, leading to marginal gains.

\subsubsection{Performance with respect to L1-I cache misses}
\begin{figure}[h]
    \centering
    \includegraphics[width=\columnwidth]{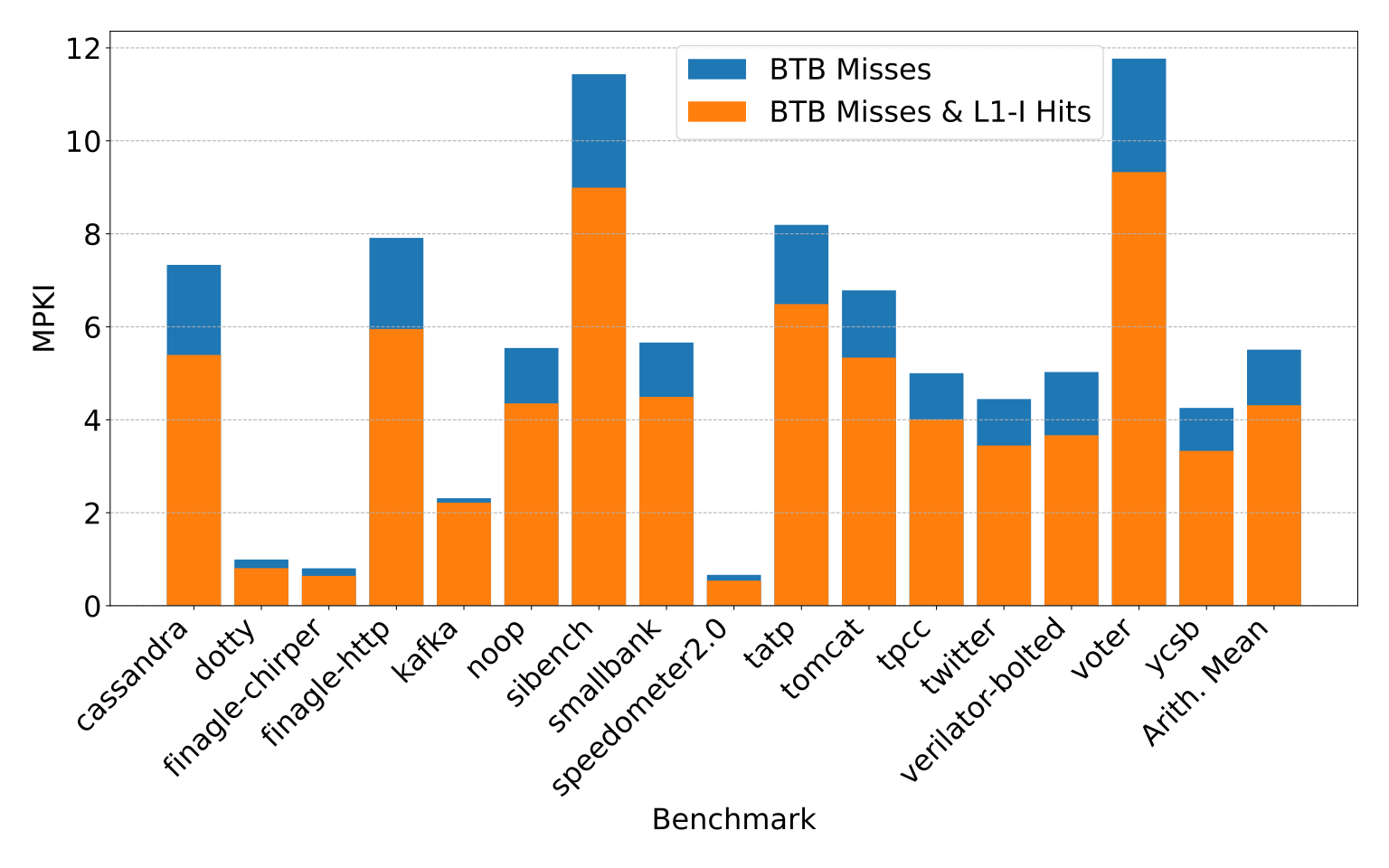}
    \caption{BTB miss with L1-I cache line hit per benchmark for 8k-entry (78KB) BTB.}
    \label{fig:L1Hit_BTB_Miss_All}
\end{figure}

Figure \ref{fig:L1Hit_BTB_Miss_All}, provides insights into all BTB miss branches. The stacked bar chart indicates a significant percentage of BTB miss branches associated with cache lines that were present in L1-I. Notably, \texttt{kafka} exhibits numerous branch cache lines experiencing L1-I cache hits but with BTB misses, in contrast to \texttt{finagle-chirper} and \texttt{speedometer2.0}. Despite this observation, the IPC gain remains minimal. This behavior can be attributed to the low occurrence of direct calls and returns, as illustrated in Figure \ref{fig:BTB_Miss_all_Type}.

\begin{figure}[h]
    \centering
    \includegraphics[width=\columnwidth]{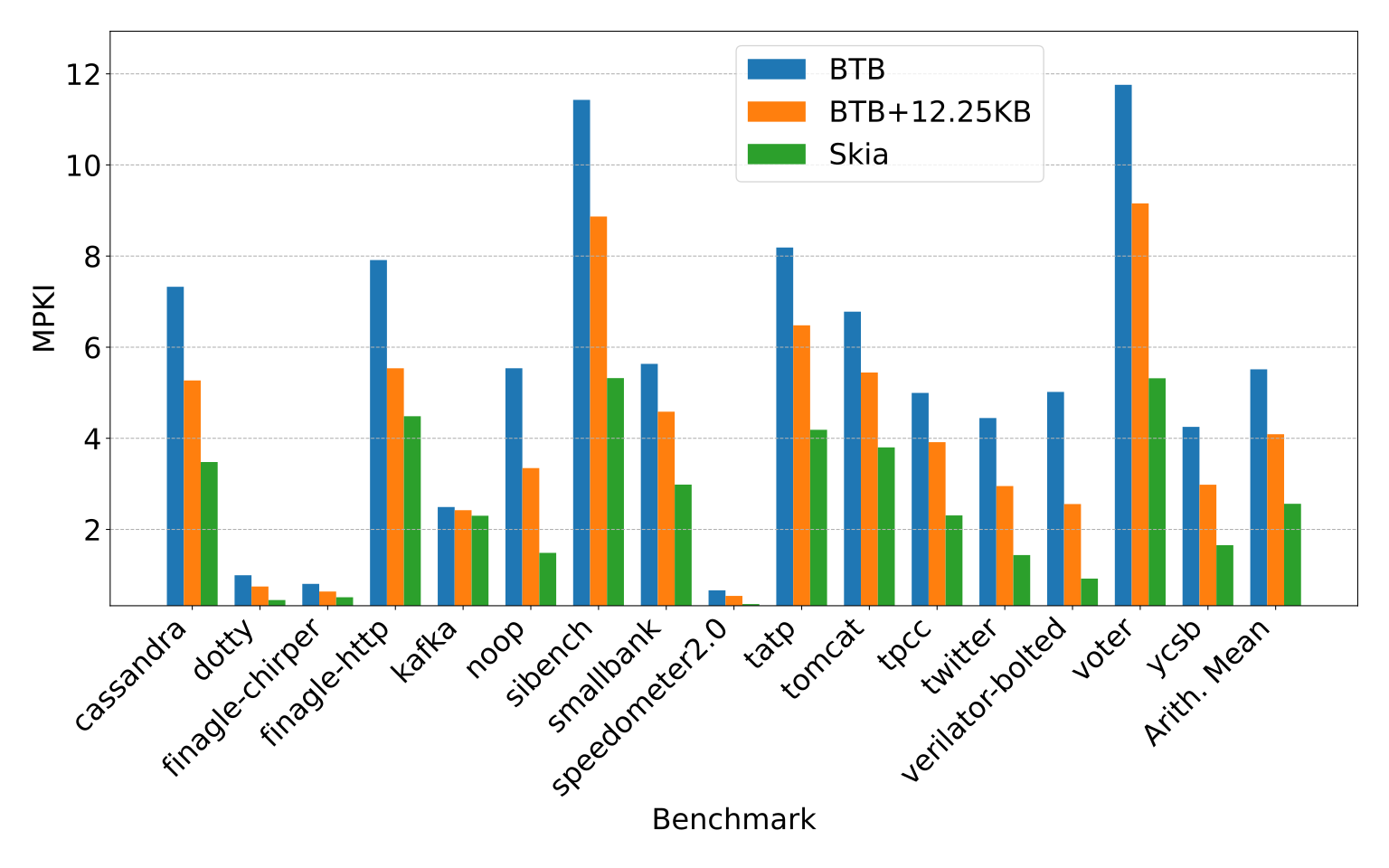}
    \caption{Overall BTB Miss MPKI comparison per benchmark for 8K-enrty (78KB) BTB.}
    \label{fig:btb_misses_mpki_reduction}
\end{figure}

\subsubsection{Skia MPKI reduction}
Figure~\ref{fig:btb_misses_mpki_reduction} shows the MPKI rate for the baseline BTB versus the same BTB with an additional 12.25KB of storage space (equal to the size of the SBB) and versus Skia with its SBB.  The figure demonstrates that Skia reduces the average BTB MPKI by $\sim$115\% when compared to the baseline BTB configuration. In contrast, allocating the same hardware budget used for the SBB to the BTB results in only a $\sim$35\% MPKI reduction.

\subsubsection{Verilator Bolted vs Pre-Bolt}
\label{sec:verilator}
While these results have been elided for brevity and to prevent confusion, we can summarize them as in general the non-bolted verilator exhibits significantly more BTB misses.  As a result, Skia improves performance significantly more than it does in the verilator-bolted shown above (10.27\%).  Given that Skia achieves significant performance gain when the application is bolted as well, this indicates that Skia provides robust gains regardless of software techniques such as BOLT.

\subsection{SBB Sensitivity Analysis}
We investigate the performance impact of scaling the sizes of U-SBB and R-SBB by varying the number of entries while maintaining a constant associativity of 4 relative to the FDIP baseline using an 8K-entry BTB.

\begin{figure}[h]
  \centering
  \includegraphics[width=\columnwidth]{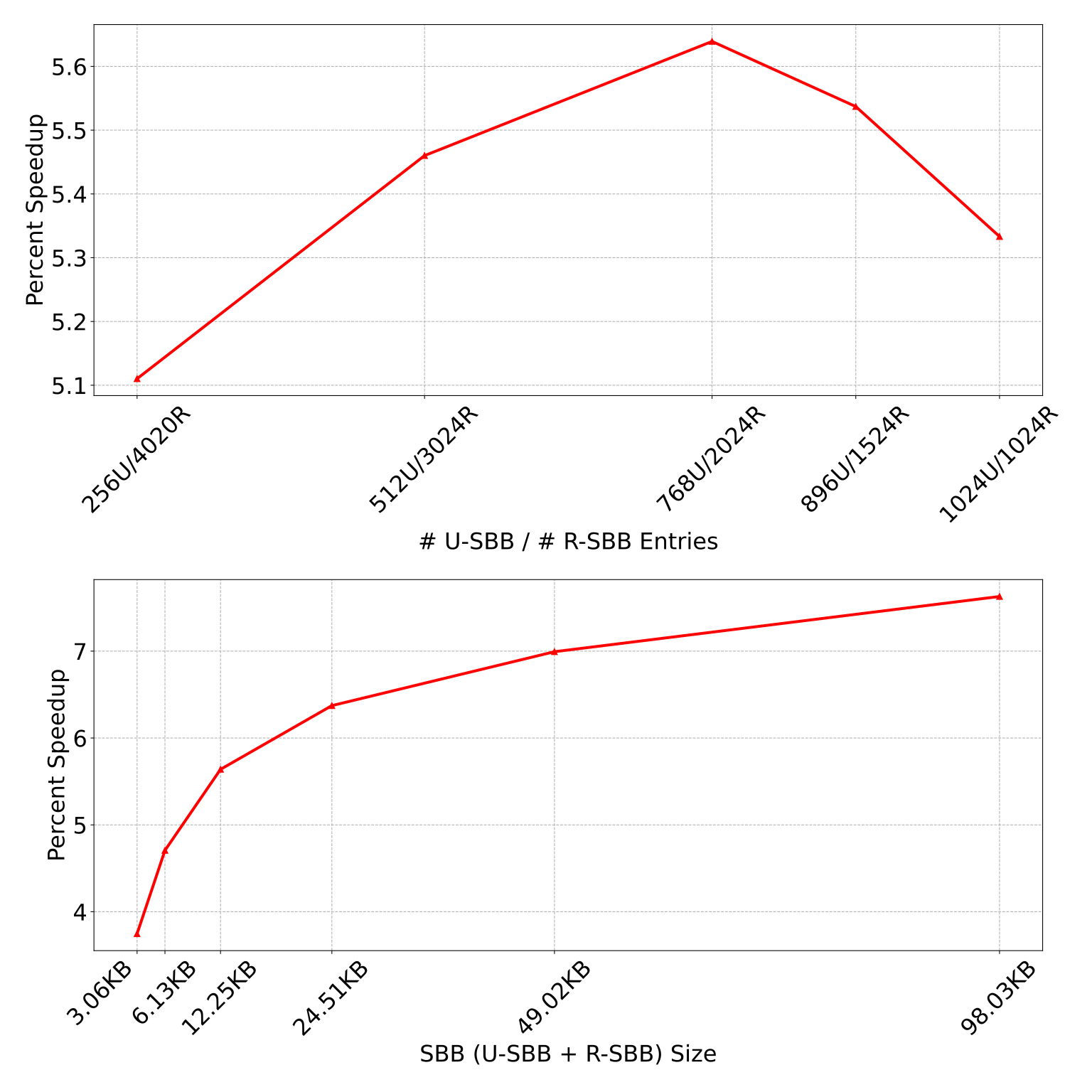}
  \caption{Top: Combination of U-SBB and R-SBB performance while keeping the size the same.  Bottom: we maintained a constant ratio of U-SBB to R-SBB entries while scaling them up to observe saturation points.}
  \label{fig:sbb_sweep}
\end{figure}

The top chart in Figure~\ref{fig:sbb_sweep} illustrates the effectiveness of combining both structures and identifies the optimal configuration while maintaining a constant state size of 12.25 KB.  The preferred setup entails allocating 768 entries to U-SBB and 2024 entries to R-SBB.  This distribution results in 7.3125 KB for U-SBB and 4.9375 KB for R-SBB, totaling 12.25 KB. 
 In the bottom chart of Figure~\ref{fig:sbb_sweep} we show how the performance is scaled if we provide more hardware budget while keeping the number of entries ratio between U-SBB and R-SBB the same.

\begin{figure}[h]
    \centering
    \includegraphics[width=\columnwidth]{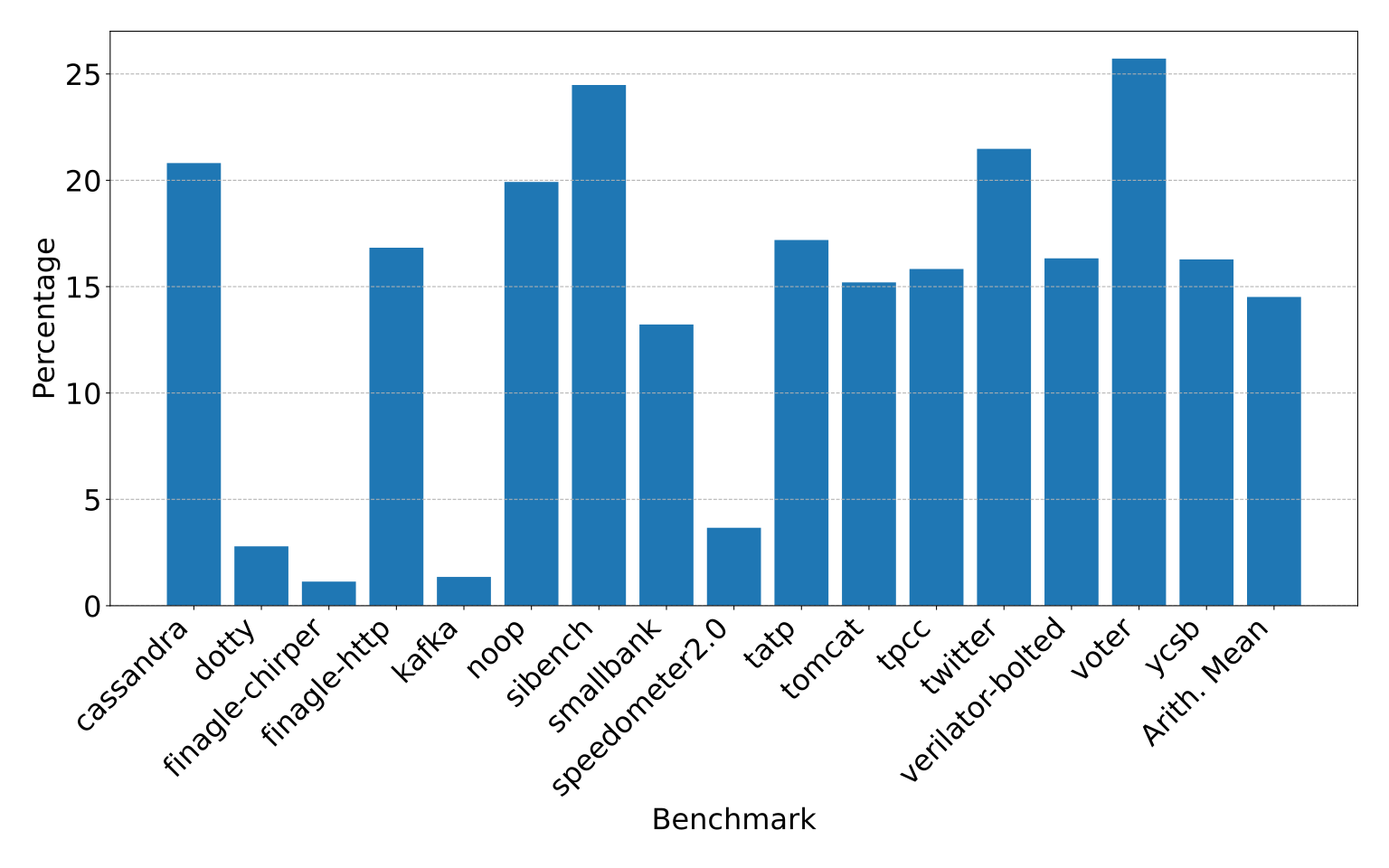}
    \caption{Reduction in decoder unit idle cycles due to Skia versus a baseline without Skia for a system with an 8K-entry (78KB) BTB.}
    \label{fig:decode_idle_cycles}
\end{figure}

\subsection{Skia and Decoder Idle Cycles}
Figure~\ref{fig:decode_idle_cycles} shows the decrease idle cycles for the core's decode stage, as a percentage versus a baseline system without Skia.  The figure highlights the percentage reduction achieved by mitigating the repair and refill costs. This improvement is possible due to the exposure of shadow branches already present in the L1-I cache. Both \textit{voter} and \textit{sibench} exhibit significant reductions at the decode stage, attributed to their high frequency of direct unconditional branches, calls, and returns, as shown in Figure~\ref{fig:BTB_Miss_all_Type}. Consequently, these two benchmarks also demonstrate substantial IPC gains.

\section{Related Work}
\label{sec:related_work}
Previous studies mitigate front-end stalls by minimizing instruction cache misses and improving BTB efficiency by introducing Hardware or Software approaches.
\subsection{Hardware-Based Approaches}

Confluence~\cite{kaynak2015confluence} introduces AirBTB, a structure that tracks the branches within cache blocks brought into the L1-I.
This information is provided to the branch predictors when a cache line is fetched to allow the front-end to continue speculating the instruction stream. AirBTB's organization allows it to track branches on a cache line basis; in particular, its design ensures that its contents are present in the L1-I, making it unlikely to retain or consistently identify cold branches present in those cache lines. 
    
Boomerang~\cite{kumar2017boomerang}, similar to Confluence, attempts to reconcile BTB inefficiencies by extracting branch information from cache lines brought into the L1-I due to a cache miss. On a BTB miss, Boomerang accesses the cache line containing the missing branch information in the L1-I or prefetches it into the L1-I and predecodes it. Any branch information in the cache line is placed in a BTB prefetch buffer until the BTB entry receives a demand request. Boomerang enables aggressive instruction stream speculation but risks polluting both the L1-I and BTB with speculative entries, especially when encountering large instruction footprints. In contrast, Skia leverages the already present cache lines in the L1-I, does not require additional access to the L1-I, and does not consume BTB bandwidth in its operation. 

Shotgun~\cite{kumar2018blasting} focuses on BTB-directed instruction fetch and BTB prefetching by dividing the BTB structure into an Unconditional BTB (U-BTB) that tracks conditional branches' targets and the spatial footprint surrounding each branch target, a Conditional BTB (C-BTB) that tracks conditional branch outcomes, and a Return Instruction Buffer (RIB) to track local control flow by recording return instructions. The authors proactively fill the BTB structures by inserting branch information predecoded from missed instruction cache lines into a BTB prefetch buffer, which migrates its entries into a corresponding BTB when it experiences a hit on one of its entries. Shotgun extends Boomerang and suffers similar challenges when faced with large instruction footprints, potentially inducing cache pollution through aggressive L1-I prefetching while not retaining cold branches in the BTB structures. 

Divide and Conquer~\cite{ansari2020divide} takes a different approach and divides its front-end prefetching mechanism to target different instruction stream behaviors. They incorporate a sequential prefetcher to identify sequential accesses and a discontinuity prefetcher to identify non-sequential accesses for prefetching into the L1-I. Cache lines prefetched by the discontinuity prefetcher are also sent to a predecoder to identify branches to place in a BTB prefetch buffer. To reduce the number of cache lookups, they track the last eight recently demanded, or prefetched, cache lines and compare incoming prefetches to them to reduce redundant lookups. The authors address variable-length ISAs and record the byte offsets of previous predecoded branches in the LLC. Again, this proposal requires prefetching into the L1-I and additional cache lookups, and while tracking recently accessed cache lines reduces redundant accesses, Skia leverages cache lines already going to the front end to identify potential BTB misses, incurring no additional L1-I traffic.

The wrong-path instruction prefetching~\cite{pierce1996wrong} technique reduces instruction cache misses by prefetching both next-line and target instructions along predicted and non-predicted paths. This approach anticipates that even mispredicted paths may become relevant soon, reducing cache misses when the processor revisits these paths. Although effective, this method slightly increases memory traffic as it brings in potentially unused instructions, which can lead to cache pollution.

The collapsing buffer~\cite{conte1995optimization} mechanism improves instruction alignment in high-issue-rate superscalar processors by realigning branches and their targets within cache blocks, ensuring a contiguous sequence of instructions reaches the decoder. This approach, using a crossbar or shifter to reorder instructions, significantly boosts decoder utilization and instruction throughput. However, it depends heavily on compiler optimizations, like code reordering, to achieve maximum efficiency. While effective, these limitations may hinder its scalability and increase resource demands in varied processor configurations.

These approaches successfully reduce the overall miss rate of the BTB but are susceptible to cold branch behavior, suffering from similar hardware overhead constraints as the BTB. Identifying infrequent cold branches requires large metadata structures that can prioritize high-impact cold BTB misses. In contrast, Skia is a low overhead mechanism that incurs no additional L1-I accesses or pollutes any structures on the critical path with speculative instructions, and leverages instruction cache lines already sent to the front-end to improve performance.

\subsection{Software-Based Approaches}
Alternatively to hardware based mechanisms, software-based profile-guided solutions have been proposed to reduce the number of BTB misses and reduce the impact of large instruction footprints.

Twig~\cite{khan2021twig} takes a software-based approach to improving BTB performance. The authors use software profiling to identify branches that result in a high number of BTB misses. Twig introduces a BTB prefetch instruction that inserts along paths with a high conditional probability of leading to a BTB miss based on a profile of the application's control flow. They further improve this approach by compressing the branch target to lower the BTB prefetch's overhead, storing key-value pairs in memory when the target cannot be compressed. The key-value pairs also encode a spatial footprint of nearby branches to be prefetched with a coalesced BTB prefetch instruction. 

Thermometer~\cite{song2022thermometer} uses software profiling to collect a trace of branch execution and then simulate an optimal replacement policy to identify a particular branch's temperature or hit-to-taken ratio. The temperature is injected as hints into the unused bits of x86 branch instructions to be passed to and stored in the BTB to inform its replacement policy, prioritizing cold branches for eviction. 

Software profiling techniques provide a low overhead solution to reduce BTB misses as the majority of collection and analysis is performed offline. However, profiling can be challenging to deploy in commercial systems as changing the underlying application may change the hints' accuracy, requiring re-profiling and re-analysis to regain performance benefits. 

\section{Conclusions}
\label{sec:conclusions}
Contemporary data center and cloud applications continue to become more complex, with increasing code footprints resulting in a high number of BTB misses. FDIP can help reduce L1-I cache misses, but it heavily depends on the contents of the BPU's tracking structures. When it encounters a BTB miss, the BPU may not identify the current instruction as a branch to FDIP, resulting in mis-speculation and decreased performance.

We observe that the vast majority, 75\%, of unidentified branches that cause BTB-misses are present in instruction cache lines that FDIP has previously fetched. We find that these branches are in the shadow of executed basic block, already present in the front-end, but are in the cache line before the branch target that brought the line into the cache or are present after a taken branch leaves the cache line.

We propose Skia, a novel shadow branch decoding technique that identifies and decodes unused bytes in cache lines already fetched by FDIP, inserting them into a Shadow Branch Buffer (SBB).

We demonstrate that Skia, with a minimal size of 12.25KB, delivers a geomean speedup of $\sim$5.7\% over an 8K-entry BTB (78KB) and $\sim$2\% versus adding an equal amount of state to the BTB, across 16 L1-I bound commercial workloads.

\balance
\bibliographystyle{plain}
\bibliography{main}

\end{document}